\documentclass[12pt,english,preprint]{emulateapj}

\slugcomment{Accepted by {\it The Astrophysical Journal}  809:80 (19pp), 2015 August 10}

 \usepackage{epsfig}
 \usepackage{epstopdf}

\shorttitle{Population Synthesis of CVs}

\shortauthors{Goliasch \& Nelson}

\makeatletter

\usepackage{babel}

\usepackage{graphicx}
\graphicspath{%
    {converted_graphics/}%
    {/}%
}
\begin{document}

\title{POPULATION SYNTHESIS OF CATACLYSMIC VARIABLES: I. INCLUSION OF DETAILED NUCLEAR EVOLUTION}

\author{Jonas Goliasch$^{1}$ and Lorne Nelson$^1$}

\affil{$^1$ Physics Department, Bishop's University, 2600 College Street, Sherbrooke, Quebec Canada J1M 1Z7\\}

\email{lnelson@ubishops.ca}

\begin{abstract}

We have carried out an extensive population synthesis study of the ensemble properties of the present-day population of cataclysmic variables (PDCVs) that takes into account the nuclear evolution of high-mass donors close to the bifurcation and dynamical instability limits. Assuming the interrupted magnetic braking paradigm, we confirm many of the general features associated with the observed CV population and find enormous diversity in their secular properties. We predict that nearly half of the non-magnetic CVs with $P_{orb} \ge 6$ hours are at least mildly evolved (i.e., $>$ 50\% of their MS turn-off age). 
Some of these systems contribute to the observed population of 	PDCVs in the period gap. We also see an enhancement by up to a factor of two in the probability of detecting CVs at the `minimum period'. This spike is quite narrow ($\approx 5$ minutes) and is attenuated because of the spectrum of WD masses and partly by the evolution of the donors. Our syntheses imply that there should be a very rapid decline in the number of ultracompact CVs (such as AM CVns). We find that between $\sim 0.05$ to 1\% of PDCVs could be UCs, and thus it is likely that the CV channel is probably not the primary contributor to the intrinsic population of UCs (especially for $P_{orb}$ < 30 minutes). Finally, a preliminary analysis of our results suggests that WDs in PDCVs experience a net gain in mass of $\lesssim 0.1 M_\odot$ as a result of high mass-transfer rates early in their evolution.

\end{abstract}


\section{\label{sec:INTRODUCTION}INTRODUCTION}

Cataclysmic variables (CVs) are a very heterogeneous class of semi-detached, interacting binaries consisting of a white dwarf (WD) that is accreting matter from a companion (donor) star that is overflowing its Roche lobe (see Patterson 1984; Warner 1995; and references therein). The recent SDSS survey has produced a wealth of new observational information (see, e.g., Szkody et al. 2011). According to the conventional model for the formation of galactic-disk CVs, the progenitors are primordial binaries for which one star is sufficiently massive and close enough to its companion star that it can engulf the companion during the giant phase of its evolution. The time interval over which this happens is governed by the nuclear timescale of the more massive star (the primary). The binary then enters a short-lived (dynamical timescale) common envelope (CE) phase of evolution during which the companion spirals inside the envelope of the giant star until it approaches the degenerate core (the nascent WD). The transfer of energy leads to the envelope being completely ejected thereby producing a binary that is most likely detached (i.e., a post common-envelope binary [PCEB]). The donor (secondary) can begin to transfer mass if the binary separation is reduced on a sufficiently short timescale through orbital angular momentum loss (AML) or if the secondary expands sufficiently due to its own nuclear evolution. Once mass transfer from the donor to the WD commences, we refer to the binary as a zero-age cataclysmic variable (ZACV). The trajectory of the matter as it undergoes Roche-lobe overflow (RLOF) largely depends on the strength of the magnetic field of the accreting WD. If the WD is non-magnetic, an accretion disk will form around the WD; but if the magnetic field is sufficiently strong, the disk can either be truncated or may not even form if the magnetic field can entrain the matter and force it to flow directly onto the WD's surface.

The subsequent evolution of the CV is either driven by orbital angular momentum losses or the nuclear evolution of the donor. According to the `canonical model' of CV evolution, donors whose masses exceed approximately 0.37 $M_\odot$ will experience some form of magnetic braking (MB). It is assumed that MB will become ineffective once the mass of the donor star has been decreased so much that its internal structure becomes completely convective. This has been referred to as the Interrupted Magnetic Braking (IMB) paradigm (see Rappaport et al. 1983 [RVJ]; Spruit \& Ritter 1983; Hameury et al. 1988; Davis et al. 2008). When MB is switched off, the donor shrinks beneath its Roche lobe leading to the cessation of mass transfer at orbital periods of $\approx 3$ hours. Gravitational radiation (GR) losses alone then drive the binary back into a semi-detached state ($\approx 2$ hours) and the system continues to evolve towards a minimum orbital period ($P_{min} \approx 80$ minutes) before returning back to higher periods.

It is important to try to infer the evolutionary history of CVs by comparing the observed ensemble properties of CVs with theoretically computed population syntheses. Although there are many subclasses of non-magnetic CVs (e.g., dwarf novae, recurrent novae) whose behaviors are governed by different physical effects, their evolutionary histories share many common features.  Cataclysmic variables are especially well-suited for the application of population synthesis (PS) techniques because they may be viewed as ``first order" systems relative to other classes of interacting binaries. Higher order systems would include, for example, low-mass X-ray binaries (LMXBs), and double neutron-star binaries.  These systems have extra dimensions of uncertainty complicating the calculation of  their formation probabilities. For example, they might have undergone two CE phases or experienced natal neutron star kicks, both of which are subject to large physical uncertainties. Thus it is important to first securely determine the relative probabilities of the various channels that lead to the formation of CVs.  The overall population of CVs is sufficiently large ($\gtrsim 1000$) that statistically significant inferences can be made if unbiased samples are available. Unfortunately, selection effects are the bane of this type of study thereby making it difficult to reach unbiased conclusions.  

The use of PS techniques as a tool for understanding the formation and evolution of all types of interacting binaries has been extensively studied.  Some of the key PS studies that have made significant contributions to the CV field include those of de Kool (1992), Kolb (1993), Politano (1996), Howell et al. (1997), Nelemans et al. (2001), Howell et al. (2001 [HNR]), Podsiadlowski et al. (2003), Politano (2004), Kolb and Willems (2005), and Willems et al. (2005, 2007).  However, the relative complexity of the physical processes and the wide range of timescales associated with them still pose considerable difficulties even with currently available computing power. The large number of dimensions of parameter space has often required that many simplifying assumptions be implemented in order to make the computations tractable.

Except for the pioneering study by Podsiadlowski et al. (2003), previous PS studies have discounted the effects of the internal chemical evolution of the donor star on the Present Day CV (PDCV) population\footnote{We also refer the reader to the work of Andronov and Pinsonneault (2004) who examined how the evolution of chemically evolved CVs is affected by various descriptions of magnetic stellar wind braking.}. In this study, we examine the effects that such evolved donor stars have on the properties of the PDCV population. Our approach is to first compute detailed evolutionary tracks using a Henyey-type code (with state-of-the-art input physics) in order to generate a grid of models that are representative of most types of non-magnetic CV evolution. We then interpolate this grid to obtain a track corresponding to a specific set of initial conditions describing a ZACV. Since the interpolation is computationally inexpensive (as opposed to the calculation of individual CV tracks), it is relatively easy to generate large populations of PDCVs from a previously synthesized group of ZACVs. 

This paper is organized as follows: In \S 2 we describe the stellar code that is used to create the CV evolution grid, the method of interpolation, and our assumptions concerning the physics of the population synthesis itself. In \S 3, we present the results of the population synthesis study and explore the observable properties of the ensemble of PDCVs; a discussion of the implications of these results can be found in \S 4. In particular, we analyze the importance of evolved donors with respect to the period gap, the minimum period, and the formation of ultracompact binaries, and we briefly comment on the possible increase in the mass of WDs in CVs. Our conclusions and plans for future work are summarized in \S 5.

\section{METHODOLOGY}

One of the most significant problems in synthesizing any population of interacting binaries relates to the large number of dimensions of parameter space that must be fully explored. In most of the previous studies, the standard approach has been to select the properties of the primordial binaries based on empirical probability formulae and then to follow the evolution of individual binaries to the present day. This is a very numerically expensive process and often requires that many simplifying assumptions be imposed or necessarily places limitations on the region of parameter space that can be investigated. 

The use of simplified codes as a means of enhancing computational speed is a trade-off that is commonly made. While reasonably accurate, the composite polytrope code used by HNR, for example, did not allow for the incorporation of chemical evolution of the donor. Our approach to synthesizing the population of ZACVs is similar to that described by HNR (Monte Carlo method) and will be outlined in \S 2.1 below. However, to obtain the PDCV population, our approach is quite different. In order to utilize a sophisticated state-of-the-art code that includes nuclear evolution as well as a more robust implementation of the input physics, our strategy is to construct a grid of highly accurate, pre-computed models. This carefully constructed grid contains representative tracks that encompass the phase space of initial conditions for ZACVs. The evolutionary tracks start from the ZACV stage and then follow the evolution over an interval of at least 12 Gyr. We can then simulate the evolution of any specific ZACV by means of interpolation, and thus greatly increase the computational efficiency without sacrificing accuracy. Once the grid has been computed, the only computational cost comes from the Monte Carlo synthesis of the ZACVs and then from the multidimensional interpolation of the evolutionary grid. A similar approach to population synthesis has been used by Podsiadlowski et al. (2003) and by Willems et al. (2005, 2007). The details of our method are discussed in \S 2.2.

\subsection{Synthesis of ZACVs}

We adopt the formulation of HNR to synthesize the Galactic-disk population of ZACVs. Using Monte Carlo techniques, we generate a set of $10^8 $ primordial binaries for each set of parameters that we investigate. Each {\it case} corresponds to a unique choice of the Initial Mass Function (IMF), mass-correlation function, primordial separation, birthrate function (BRF), and other physical inputs (e.g., CE parameters and magnetic braking). We assume an approximately solar metallicity (appropriate for the disk) throughout the calculation. We also ignore the formation of multiple star systems. 

\subsubsection{Primordial Binaries}

Based on a particular IMF, the initial mass of the primary ($M_{10}$) is the first parameter to be chosen. The primary is always taken to be more massive than the secondary ($M_{10} > M_{20}$). Our preferred choice of the IMF is based on the work of Miller \& Scalo (1979) for which the masses obey a log-normal distribution (half-Gaussian in $\log M$). We also consider the Salpeter (1955) IMF with a mass index of 2.35. Rather than relying on an analytic approximation for the Miller \& Scalo IMF, we use the Von Neumann approach (see, e.g., Sobol 1994) to carry out the Monte Carlo analysis. This has the advantage of faithfully reproducing the parent distribution at both low and high masses to arbitrary accuracy.
In particular, we note that this method is superior to the analytic formula devised by Eggleton et al. (1989; and used by HNR) because the latter underestimates the Miller-Scalo results by up to 25\% for the lowest-mass stars.\footnote{The standard Monte-Carlo approach relies on an analytic transformation of a uniform random variable to a random variable $\zeta$ that obeys the desired probability density function $p(x)$ with upper bound $K_{0}$ (i.e., $p(x)\leq K_{0})$ on the interval $a\leq x\leq b$. In lieu of such an analytic transformation, the Von Neumann Method employs the following two-step acceptance-rejection algorithm:
\break
1) Generate two uniformly distributed random numbers $\chi'$ and $\chi''$ on [0,1] and construct the point $\Gamma(\mu',\mu'')$ with coordinates $\mu'=a+\chi'(b-a)$, and $\mu''=\chi''\, K_{0}$.\hfil
\break
2) If the point $\Gamma$ lies under the curve (i.e. $\Gamma\leq p(x)$), then set $\zeta=\mu'$; otherwise reject $\Gamma$. \hfil
\break
By repeating this algorithm continuously, the accepted values of $\zeta$ will follow the desired probability density distribution.
}

The initial masses of the secondary are chosen for several different correlation functions. Empirical evidence suggests that there is a weak correlation between the masses of the two components, but the correlation may not be a monotonic function of the primary mass. The probability distribution that we adopt was first suggested by Abt \& Levy (1978) and is expressed as a power law in terms of the mass ratio $q$ ($q \equiv M_{{2}}/M_{{1}}$). We take
\begin{equation}
\frac{dN}{dq} \propto q^{\xi}
\end{equation}
\noindent where $\xi$ is the power law index.  For our standard case (Case 1), $\xi=\frac{1}{4}$ (see Duquennoy and Mayor 1991). To explore the dependence of our results on the correlation, we also generate populations with: (i) $\xi=1$; (ii) no correlation between primary and secondary masses; and, (iii) perfect correlation between primary and secondary masses (i.e., $\xi=\infty$ and thus $q = 1$).

The orbital period distribution of the primordial binaries is commonly chosen to be either a uniform distribution in the logarithm of the orbital period ($P_{orb}$) or of the orbital separation. We follow the suggestion of Abt \& Levy (1978) and adopt an orbital period distribution that is uniform in $\log P_{orb}$ for periods in the range of 1 day to $10^6$ years. 

Once the properties of the primordial binaries are enumerated ($M_{10}, M_{20}, P_{orb}$), only systems that were capable of becoming ZACVs are considered. Primordial binaries are rejected if: (1) the separation was so large that the primary could never expand sufficiently to fill its Roche lobe; (2) the separation was so small that the primary was immediately overfilling its Roche lobe presumably leading to a merger; (3) the separation was insufficient to allow the primary to expand to a point where it could form a degenerate core (nascent WD) before making contact with its Roche lobe; (4) the mass of the secondary (donor) star was so low that it was not on the main sequence (i.e., it was a brown dwarf\footnote{The existence of brown dwarfs (BDs) as CV companions has been studied by, for example, Politano (2004).}). We took the minimum mass to be $0.08\, M_{\odot}$ even though our detailed models show that brown dwarfs can have masses as small as $0.072\, M_{\odot}$. Note that these criteria for rejecting primordial binaries and for determining the mass of the white dwarf companion ($M_{WD}$) are the same as those adopted by HNR (except that HNR set the cut-off mass for the secondary to be $0.09\, M_{\odot}$); the details are discussed more extensively in that paper. 

\subsubsection{Common Envelope Evolution}

If all of the criteria listed in \S 2.1.1 are met, the primordial binary is assumed to undergo a phase of CE evolution. We have checked ({\it a posteriori}) that the vast majority of systems will indeed undergo unstable mass transfer based on the criterion given by Hurley et al. (2002). According to first principles energy arguments describing CE evolution, frictional heating provides the requisite energy necessary for the ejection of the primary's envelope.  The actual hydrodynamics of the CE phase is difficult to model (see, e.g., Taam et al. 1978; Taam \& Sandquist 1998), and will not be discussed here\footnote{It has been claimed that the CE phase can lead to an expansion of the orbit when considering angular momentum rather than energy conservation (see, e.g., Nelemans \& Tout 2005, and references therein). However, this approach has received considerable criticism (Woods et al. 2012).}. 

To determine the properties of the PCEB, we use the same procedure described by HNR. Essentially we assume that the energy needed to unbind the envelope of the primary is some fraction of the change in the orbital energy.  Thus $E_{bind} = \epsilon \Delta E_{orb}$, where $\epsilon$ is the dimensionless CE efficiency factor (normally $< 1$).  Following HNR, we take    
\begin{equation}
\epsilon \, \frac{ GM_{2}}{2}\left(\frac{M_{core}}{a_{f}}-\frac{M_{10}}{a_{0}}\right)=\frac{G\left(M_{env}+3M_{core}\right)M_{env}}{R_{L,1}}\label{eq:Common Envelope}
\end{equation}
\noindent where the primordial orbital separation is $a_{0}$, the final orbital separation  is $a_{f}$, and $R_{L,1}$ is the primary's Roche lobe radius.  Note that the WD mass is taken to be equal to $M_{core}$, and $M_{env}$ represents the mass of the primary's envelope ($= M_{10}-M_{core}$). This formulation is similar to one that is commonly used (e.g., Webbink, 1984; Willems et al., 2005), except that the dimensionless structure parameter determining the binding energy of the envelope to the core (commonly referred to as $\lambda$) has not been explicitly calculated. Equation (2) does constrain the range of possible values of $\lambda$ to be between 1/3 and 1.  Usually $M_{env} >> M_{core}$, thereby constraining $\lambda$ to be $\lesssim 1$. Other formulations for energy conversion have been proposed (see, e.g., Iben and Livio 1993; Taam and Sandquist 2000; Tauris and Dewi 2001; Politano and Weiler 2007; De Marco et al. 2011) and may be more useful; but we caution that there are very large uncertainties in any of these formulations.  We plan to investigate these other possibilities in the future.

\subsubsection{Post Common Envelope Evolution}

If the Roche lobe is found to be located within the donor's surface after the CE phase, it is presumed that a merger would have occurred and thus the system is rejected.  If the PCEB is in a detached state, the system can become a ZACV if the orbital separation decreases sufficiently due to AMLs and/or the donor expands on its nuclear timescale and overflows its Roche lobe.  If the PCEB's orbital separation is quite large, then the secondary can be significantly evolved by the time the system reaches a semi-detached state. While systems with chemically evolved secondaries are technically CVs, their evolution can differ markedly from the canonical one described in \S 1. If the nuclear timescale of the secondary is less than (or approximately equal to) the mass-loss timescale, $P_{orb}$ can increase as the secondary turns off the main sequence (and may even form a degenerate core). The dividing line between these two radically different evolutionary behaviors is known as the bifurcation limit (see, e.g., Pylyser \& Savonije 1988; Nelson et al. 2004a [NDM]). For systems above this limit, mass loss occurs when the donor is in the Hertzsprung gap or on the red giant branch and thus can form a degenerate helium core around which hydrogen shell-burning occurs (i.e., stars above the Sch\"{o}nberg-Chandrasekhar Limit). The donor expands generally on a nuclear timescale and contracts rapidly after losing most of its envelope mass leaving a double-degenerate, detached binary. In this paper, only systems that evolve below the bifurcation limit are considered (the others are ignored). The rejected systems would evolve as long-period CVs ($P_{orb} \gtrsim 1$ day) and could potentially form double-degenerates.  This possibility will be examined in a subsequent paper.

When the secondary starts RLOF, the binary could be subjected to a dynamical instability (see, e.g., Paczy\'nski 1967; Hjellming \& Webbink 1987; Soberman et al. 1997).  Dynamical instabilities occur when mass transfer causes the radius of the Roche lobe to move inwards relative to the surface of the star on a timescale much shorter than its thermal adjustment timescale (Kelvin-Helmholtz [KH] time). This precipitates a runaway situation and very often leads to the merger of the two stars.  Using our grid of evolutionary tracks and by computing additional tracks for which dynamical instabilities can occur, we are able to delineate stable from unstable initial conditions. A detailed examination of this issue can be found in Nelson \& Goliasch (2015).

Finally, the ages of the systems that could ultimately undergo stable mass transfer are checked. If the time required for the formation of the ZACV exceeds the age of the galaxy, taken to be $10 ^{10}$ yr, the system is rejected (see Section 2.2.4). Once this final check is complete, the surviving binaries constitute the final set of all incipient ZACVs for a particular case (i.e., for a specific set of PS parameters). The cases that we investigate are listed in Table 1.

\subsection{Synthesis of PDCVs}

For each of the PS cases, we construct a dataset describing the initial conditions for each member of the ensemble of ZACVs. These properties include $M_{donor} (\equiv M_2)$, $M_{WD}$, and the age of the system.  Each `triple' constitutes the minimal set necessary to determine all of the other properties of the ZACV.  For example, they determine the evolutionary state of the donor  (denoted by its central hydrogen content, $X_c$). 

Based on the closeness of the orbits, tidal effects enforce circularization of the orbit and synchronization of the donor.  Angular momentum losses include (when applicable) both the effects of GR and magnetic stellar wind (MSW) braking.  AML due to GR was calculated using the Landau \& Lifshitz (1962) quadrupole formula, while we use the RVJ parameterization of the Verbunt \& Zwaan (1981 [VZ]) magnetic braking law (i.e., $\dot J \propto R^{\gamma}$ where $\gamma$ is a dimensionless number between 2 and 4).  For the present investigation we set $\gamma = 4$ since this value reproduces the observed period gap reasonably well and matches the original VZ law.  We also self-consistently take into account the moment of inertia of the donor when implementing this formula (see NDM for more details). The interior structure of the donor is continuously monitored to check whether it has a radiative core (and convective envelope); whenever the donor becomes fully convective, MB is set equal to zero.  

Since mass-transfer is assumed to be completely non-conservative, we set the mass-capture fraction (${\beta \equiv \dot M_{WD} / {\vert \dot M_{donor} \vert}}$) equal to zero, and further assume that the matter lost from the system carries away a specific angular momentum equal to that of the WD (fast Jeans' mode).  We note that this assumption seems to be well warranted when CVs experience mass-transfer rates of $\lesssim 10^{-9} M_\odot$/yr (see, e.g., Prialnik \& Kovetz 1995).  At this rate, the matter transferred to the WD is not permanently accreted but rather undergoes a series of thermonuclear explosions (Classical Novae).  It is certainly possible that the WDs are even eroded during these events ($\beta < 0$).  However, during other phases of evolution 
(e.g., thermal timescale mass transfer) it may be possible for the mass of the WD to grow (i.e., $0 < \beta \leq 1$). These systems would resemble supersoft X-ray sources (van den Heuvel et al. 1992). Finally, we  do not include the effects of X-ray irradiation on the donor (the effects are expected to be small for mass transfer rates of $\lesssim 10^{-9} M_\odot$/yr).

\subsubsection{Stellar Evolution Code}

Our stellar evolution code has been extensively tested and used for similar types of mass-loss calculations (see, e.g., Nelson et al. 1985; Dorman, Nelson \& Chau 1989, and NDM for details). The code is based on the Henyey method for which mass is treated as an independent (Lagrangian) variable. The equation of state (EOS) was constructed  using the SCVH EOS (Saumon, Chabrier, \& Van Horn 1995) as its core. The independent variables are $P_{g}$ (gas pressure) and $T$ (temperature).  An additional data table was generated to complement the SCVH tables that incorporated radiation pressure, molecular hydrogen formation, and Coulombic effects (in addition to arbitrary electron degeneracy).  The two sets of tables were spliced together and then finely interpolated (Akima algorithm\footnote{The Akima (1972) algorithm is a cubic-spline interpolation that greatly reduces oscillatory overshoot and thus gives the interpolation a more physically faithful representation.}) to obtain the final set of EOS tables.  For a detailed description of the SCVH EOS, the EOS used for the extension of the tables, as well as the interpolation and consolidation process, see Maisonneuve (2007). The prescription for mass loss due to stellar winds is that due to Reimers (1975) with $\eta = 1$. 

Radiative and conductive energy transfer was calculated using the updated OPAL radiative opacities (Iglesias \& Rogers, 1996) in conjunction with the low-temperature opacities of Alexander \& Ferguson (1994), and the Hubbard \& Lampe (1969) conductive opacities (including Canuto's relativistic corrections).  The details of the splicing of these three regimes are discussed in detail by NDM.

An additional refinement to the code concerns the calculation of the outer boundary conditions needed by the Henyey solver to compute the interior structure of the donor star. We set the matching point (i.e., the `fitting point') at a depth enclosing $99\%$ of the total stellar mass. It is at this point that the envelope integration matches the interior solution.  We pre-computed the outer boundary condition tables for an extremely wide range of (observable) stellar properties.  The main assumption is that no nuclear or gravothermal energy generation occurs in the envelope. The tables were generated by integrating the stellar structure equations from the bottom of the atmosphere down to the fitting point for a range of values of the variables $L$ (luminosity) and $T_{eff}$ (effective temperature), as well as the total stellar mass. The initial conditions for this envelope integration were interpolated from the results of the atmospheric integrations provided by the Phoenix library (see, e.g., Allard et al. 1997, Allard, Hauschildt, \& Schweitzer 2000).  The value of the (solar) metallicity was taken to be $Z=0.01732$.  A more detailed description of the atmospheric and envelope integrations that produced the outer boundary condition tables can be found in Maisonneuve (2007). 

For the grid of tracks calculated in this paper, we do not include the effects of rotation and distortion in modeling the donor star. We also ignored the effects of any WD magnetic fields. 

\subsubsection{Grid Interpolation}

Our pre-computed grid is composed of over 300 tracks each containing $10^5$ to $10^6$ models. This grid of tracks has three dimensions corresponding to the triple that defines a specific ZACV: donor mass, WD mass, and age of the donor at the onset of mass transfer ($t_{ZACV}$). There are 31 initial masses for the secondary ($M_{20}= 0.08M_{\odot}$, $0.09M_{\odot}$, $0.10M_{\odot}$, $0.15M_{\odot}$, $0.20M_{\odot}$, $...,\,2.8M_{\odot}$, in increments of $0.1M_{\odot}$) and 6 WD mass grid points ($M_{WD}=0.2M_{\odot}$, $0.4M_{\odot}$, $0.6M_{\odot}$, $1.0M_{\odot}$, $1.4M_{\odot})$. The values of $t_{ZACV}$ can be thought of as corresponding to different stages of chemical evolution of the donor and range from ZAMS up to the bifurcation limit (i.e., $t_{ZAMS} \leq t_{ZACV} \leq t_{bif}$). Letting $X_{c, \, ZACV}$ denote the hydrogen mass fraction remaining at the donor's center at the time of first contact (ZACV), our intervals for the values of $t_{ZACV}$ typically coincide with $t_{ZAMS}$, $t(X_{c, \, ZACV}$=0.7), $t(X_{c, \, ZACV}$=0.6), $t(X_{c, \, ZACV}$=0.5), $t(X_{c, \, ZACV}$=0.4), $t(X_{c, \, ZACV}$=0.2), and $t(X_{c, \, ZACV}$=0.1).\footnote{We assumed that all donor stars have reached ZAMS. Since the contraction of protostars on the Hayashi track is relatively fast, our assumption is justified for all binaries except for those with the most extreme mass ratio (massive primaries which quickly become giants and very low-mass stars that might take $10^8$ years to reach the main sequence). Even then, their radii are not significantly different from the corresponding ZAMS values. For this reason, we assume that the evolutionary track of a CV with a pre-ZAMS secondary at the time of contact is reasonably well approximated by a system with a ZAMS secondary.}


\begin{deluxetable}{lcccc}

\tablecaption{\label{tab:PopSynParameters} Population Synthesis Parameters}

\tabletypesize{\scriptsize} \tablewidth{0pt}

\tablehead{ \colhead{Case} & \colhead{CE Eff.} & \colhead{$ dN/dq \propto q^{\xi} $} & \colhead{$BRF \propto e^{ -t/\tau }$}& \colhead{IMF} \\
\colhead{\#}& \colhead{$ \epsilon $}& \colhead{$ \xi $} & \colhead{$\tau$}& \colhead{} }
\startdata
& & & & \\
1 & 1 &  ${1/4}$ & $\infty$ & MS \\
2 & 1 &  ${1  }$  & $\infty$ & MS\\
3 & 1 &  uncorr & $\infty$ & MS\\
4 & 1 &  ${1/4}$   & $3.0 $ Gyr & MS\\
5 & ${1/3}$ &  ${1/4}$  & $\infty$ & MS\\
6 & ${1/3}$ &  ${1  }$  & $\infty$ & MS\\
7 & ${1/3}$ &  uncorr  & $\infty$ & MS\\
8 & ${0.01}$ &  ${1/4}$  & $\infty$ & MS\\
9 & ${1}$ &  $\infty$  & $\infty$ & MS\\
10 & 1 &  ${1/4}$  & $\infty$ & Salpeter\\
11 & 1 &  ${1/4}$  & 0 & MS\\
& & & & \\

\enddata

\end{deluxetable}

The grid is quite coarse in terms of the spacing in $M_{WD}$.  The reason for this is that the properties of the CV evolutions are least sensitive to this dimension of parameter space.  This is largely due to the fact that Roche geometry requires that the orbital period be nearly independent of the mass of the WD. Many additional tracks were added to the grid near the bifurcation limit ($t \simeq t_{bif}$) because of the sensitivity of the evolution to a relatively small subset of initial conditions. Even with these improvements, we could not fully sample systems that were born in this relatively limited volume of (initial-condition) phase space.  Thus we underestimate the number of CVs that can evolve to ultra-short orbital periods (defined to be $<$ 60 minutes). In fact, we do not have any tracks that evolve to periods of less than 25 minutes because of the extreme fine-tuning that would be required to generate those tracks.  Based on our assumed AML mechanisms and other population parameters, we find that extremely few ZACVs fall into this category. 

With respect to dynamical instabilities, we reject any CVs that experience either an immediate or delayed (latent) dynamical instability (see Podsiadlowski et al. 2003 for more details).  We also reject any systems for which $\dot M $ exceeds $\sim 10^{-6} M_\odot$/yr.  Although this value is roughly an order of magnitude lower than the Eddington limit, it is sufficiently large to cause the WD to enter a red-giant phase resulting in the presumed merger of the two components.  Since we assume mass transfer to be fully non-conservative, and since this will be increasingly less accurate as $\dot M$ exceeds $10^{-8} M_\odot$/yr, our evolutionary tracks will be less reliable for phases of high $ \dot M$.  However, most CVs spend the vast majority of their lives evolving at rates that don't exceed this value (see \S 3).

\subsubsection{Birth Rate Function}

The rate of binary formation throughout the history of the Galaxy must be established before inferring the population properties of PDCVs in the disk.  For simplicity, we assume that the stellar IMF is constant in time (i.e., $\partial N/\partial m\equiv \Phi (m,t)=\Phi (m))$, and we further assume that 50{\%} of all stars born in the disk are members of binary systems (this latter fraction can easily be adjusted). In addition, we assume that the (binary) birthrate function (BRF) is independent of mass (i.e., $\Psi \left( {m,t} \right)=\Psi \left( t \right))$. 

The most physically representative BRFs that we consider are: (i) a constant birthrate, $\Psi \left( t \right)=C$ (i.e., equal numbers of stars are born in equal intervals of time); (ii) a ``starburst'' Dirac-delta function (all stars are born simultaneously when the Galaxy was formed); and, (iii) an exponentially decaying birthrate of $\Psi (t)=K{\mkern 1mu}e^{-t/\tau }\mbox{\thinspace }\mbox{with}\;\tau =3.0$ Gyr. For our standard case (\#1), we assume that the BRF is constant.

To obtain absolute estimates of the number of PDCVs, we normalize the results to a specific birthrate of WDs in the disk of the Galaxy. Based on the results of Holberg et al. (2008) who claim a local WD density of $5 \times 10^{\mathrm{-3\thinspace }}$pc$^{\mathrm{-3}}$, and Vennes et al. (1997) who find that the density of DA WDs should be about twice that large (assuming a constant BRF), we adopt a WD formation rate of 0.4 WD/yr. This choice is also in reasonable accord with the estimate by Smith (1997), but we caution that the rate could easily be different by more than a factor of 2. The CV birthrate can be scaled to accommodate any change in the WD formation rate. For the other (non-constant) BRFs that we consider, we normalize the results to yield the same total number of WDs that would have been born in the Galaxy.

\subsubsection{Statistical Methodology}

Since the physical state of a particular CV at any time\footnote{ We define $t=$ 0 to be contemporaneous with the formation of the Galactic disk.} can be fully described by the set of all values corresponding to its $k$ properties (i.e., \textbf{\textit{Y}}$_{k}$ = \textbraceleft $P_{orb}$, $M_{donor}$, \textellipsis \textbraceright), it is straightforward to construct a statistical picture of the ensemble properties of PDCVs if those properties have been calculated for every CV up to $t=t_{gal} =10^{10}$yr. The basic approach is discussed below and our optimization of it is described in \S 2.2.5. 

The first step begins with the selection of a sample of $N$ primordial binaries based on the probability rules described in the previous sections. Let an age then be assigned to each primordial binary based on the assumed BRF using a Monte-Carlo approach ($t_{primordial}$). Systems that successfully reach the PCEB phase are subsequently evolved until they begin RLOF (i.e., become ZACVs). The amount of time that it takes them to reach the ZACV phase \textit{after} the formation of the primordial binary is designated as $t_{ZACV}$. If $t_{primordial}+t_{ZACV\thinspace }\ge t_{gal}$, then the system is discarded. For the $N_{ZACV}$ systems that survive, each ZACV is defined by a unique `triple' at the onset of mass loss (i.e., $M_{donor}$, $M_{WD}$, $X_{c,\, ZACV})$. By interpolating the grid of evolutionary models that encloses this triple, it is possible to evaluate \textbf{\textit{Y}}$_{k}$ for each present-day CV at an evolutionary age of $t_{ev\thinspace }= t_{gal\thinspace }$-- $t_{primordial}$ -- $t_{ZACV}$. Although the physical variables describing the properties of PDCVs in \textbf{\textit{Y}}$_{k}$ are continuous, it is convenient to discretize them using a finely-spaced set of intervals (bins) such that each bin is labeled with the running index $m_{k}$ (i.e., $m_{k\thinspace }=$ 1, \textellipsis , $n_{k})$.  Note that $n_{k\thinspace }$is the total number of intervals\footnote{ In this paper we take $n_{k\thinspace }=$ 1000 regardless of the property.} assigned to the property $k$. By counting the number of systems sharing the same interval, $m_k$, it is straightforward to create 1-D and 2-D (after subdivision) histograms describing the frequency distribution of any of the physical properties of the ensemble of PDCVs. 

The final step in the process is to scale the counts so that the total number of WDs that would have been born in the Galaxy is $4 \times 10^{\mathrm{9}}$. After the scaling is performed, the frequency corresponding to any property $k$ residing in the interval $m_{k}$ is equal to the number of PDCVs that have a value of that property falling within the range of that interval (let the frequency array be defined as \textbf{\textit{Z}}$_{{k},{m_k}}$). Thus the sum of the array for a specific value of $k$ is just equal to the total number of PDCVs in the disk of the Galaxy (i.e., $\sum\limits_{m_{k} } {\textbf{\textit{Z}}_{{k},{m_k}} } = N_{PDCV}$).

\subsubsection{Statistical Optimization}

We greatly optimize the quality of our statistical inference for the counts contained within \textbf{\textit{Z}}$_{{k},{m_k}}$ by incorporating all admissible values of $t_{primordial}$ in the calculation. Rather than evaluating the properties of a CV based on a single stellar model that has an age $t_{primordial}$ in an evolutionary sequence, we evaluate the properties for all possible stellar models corresponding to the chosen triple. Although an identical approach was employed by HNR, our algorithm is much more efficient because with our pre-computed grid of CV evolutionary tracks we only need to carry out interpolations in order to evaluate the properties. Once one of the equi-probable triples is computed (a relatively minor numerical expense), every one of the models corresponding to PDCVs with evolutionary ages between $t_{ev} = 0$ and $t_{ev} = t_{max}$ ($\equiv t_{gal} - t_{ZACV}$) is used to construct the inference. This range of values for $t_{ev}$ corresponds to $t_{primordial} = t_{gal} - t_{ZACV}$ (i.e., for a system just becoming a ZACV [$t_{ev} = 0$] at the current epoch) and to $t_{primordial} = 0$ (i.e., a primordial binary that is born immediately after the formation of the disk), respectively. The BRF evaluated at $t_{primordial}$ determines the relative probabilities for the formation of the primordial binaries from which all of these possible PDCVs would have evolved.  

Specifically, for every triple describing the initial conditions of each ZACV, we calculate a {\it statistical} weight, $\Delta Q_{ij}$, for each time step ($\Delta t_{ij})$ in the evolutionary sequence. The index $i$ denotes the particular step (model) in the time sequence of the CV's evolutionary track and the index $j$ identifies the specific triple. Thus each value of $i$ for a given sequence ($j)$ corresponds to a unique evolutionary age, $t_{ev}$, as measured from the first model of the sequence (the incipient ZACV). Thus for a specific triple ($j$), $t_{ev} = \sum\limits_i {\Delta {t_{ij}}}$. Given that the duration ($\delta t)$ that a CV spends in a particular phase of its evolution is directly proportional to the probability that the CV will be observed having those same properties (defined by the values of $m_{k}$), we require that $\Delta Q_{ij}$ be linearly proportional to $\Delta t_{ij}$. Also, because $\Delta Q_{ij}$ is completely independent of $k$, its statistical weight must be applied equally to all $k$ properties that fall in the interval defined by $m_{k}$. Thus for the entire PDCV population, the {\it cumulative} weight, $W_{m_{k}}$, assigned to interval $m_{k}$ will simply be equal to the sum of all of the $\Delta Q_{ij}$s that fall within that interval.
 
We further impose the condition that the sum of all of the statistical weights ($Q_{tot})$ for every ZACV triple in our synthesis and over all evolutionary models ($0 \leq t_{ev} \leq t_{max}$) be equal to the number of PDCVs in the disk of the Galaxy. Thus we require that
\begin{equation}
{Q_{tot}} \equiv \sum\limits_{i,j} {\Delta {Q_{ij}}}  = {N_{PDCV}} \quad .
\end{equation}
This prescription guarantees that the probability of finding a PDCV with a value of the observable property $k$ within the range defined by $m_{k}$ is
\begin{equation}
{P_{{m_k}}} = \frac{{{W_{{m_k}}}}}{{{Q_{tot}}}} \quad .
\end{equation}

The value of $\Delta Q_{ij}$ must also be proportional to the probability that the primordial binary was born at an age of $t_{primordial} = t_{gal} - t_{ZACV} - t_{ev}$. In order to satisfy these conditions and to guarantee the correct normalization, the statistical weights are calculated as:
\begin{equation}
\Delta {Q_{ij}} = \frac{{{\rm{ }}{\Psi _{ij}}\left( {{t_{gal}}{\rm{ }} - {\rm{ }}{t_{ZACV}}{\rm{ }} - {\rm{ }}{t_{ev}}} \right)}}{N}{\rm{ }} \ \Delta {t_{ij}}
\end{equation}
where $N$ is the number of primordial binaries initially considered in the synthesis and $\Psi_{ij}$ is the BRF\footnote{ We use the subscripts \textit{i,j} because the argument of $\Psi $ depends on both of these indices. Also note that we define $\Psi $(\textit{t\textless }0) $=$ 0.}. Thus for each system $j$, the optimization allows the BRF to be evaluated over a range of evolutionary ages extending from $t_{ev}=$ 0 to $t_{ev}= t_{max}$. The sum $\sum\limits_i {{\Psi _{ij}}} \Delta {t_{ij}}$ yields the number of PDCVs that would have formed if all of the primordial binaries evolved so as to produce ZACVs with the corresponding triple $j$. The ensemble average number of PDCVs can be determined by calculating the expectation value over all primordial binaries used in the synthesis. Therefore 
\begin{equation}
{Q_{tot}}{\rm{ }} = {\rm{  }}\frac{{\sum\limits_{j = 1}^{{N_{ZACV}}} {\left( {\sum\limits_i {{\Psi _{ij}}{\rm{ }}\Delta {t_{ij}}} } \right)} }}{N}{\rm{ }} = {\rm{ }}{N_{PDCV}}  \quad .
\end{equation}
This algorithm is completely general.

\section{Population Synthesis Results}

The orbital period and mass-accretion rate are two of the most important variables because they can often be observationally inferred (see Patterson 1984). Orbital periods are typically measured to within a precision of better than a few minutes. The instantaneous mass-transfer rate can sometimes be inferred from the accretion luminosity, but it is not necessarily representative of the long-term secular average rates. Other quantities that can sometimes be inferred include the spectral type of the donor ($T_{eff}$), the gravitational acceleration at its surface, its mass, and the mass ratio of the binary components, $q$.  


\begin{figure} 
\centering
\includegraphics[scale=0.55]{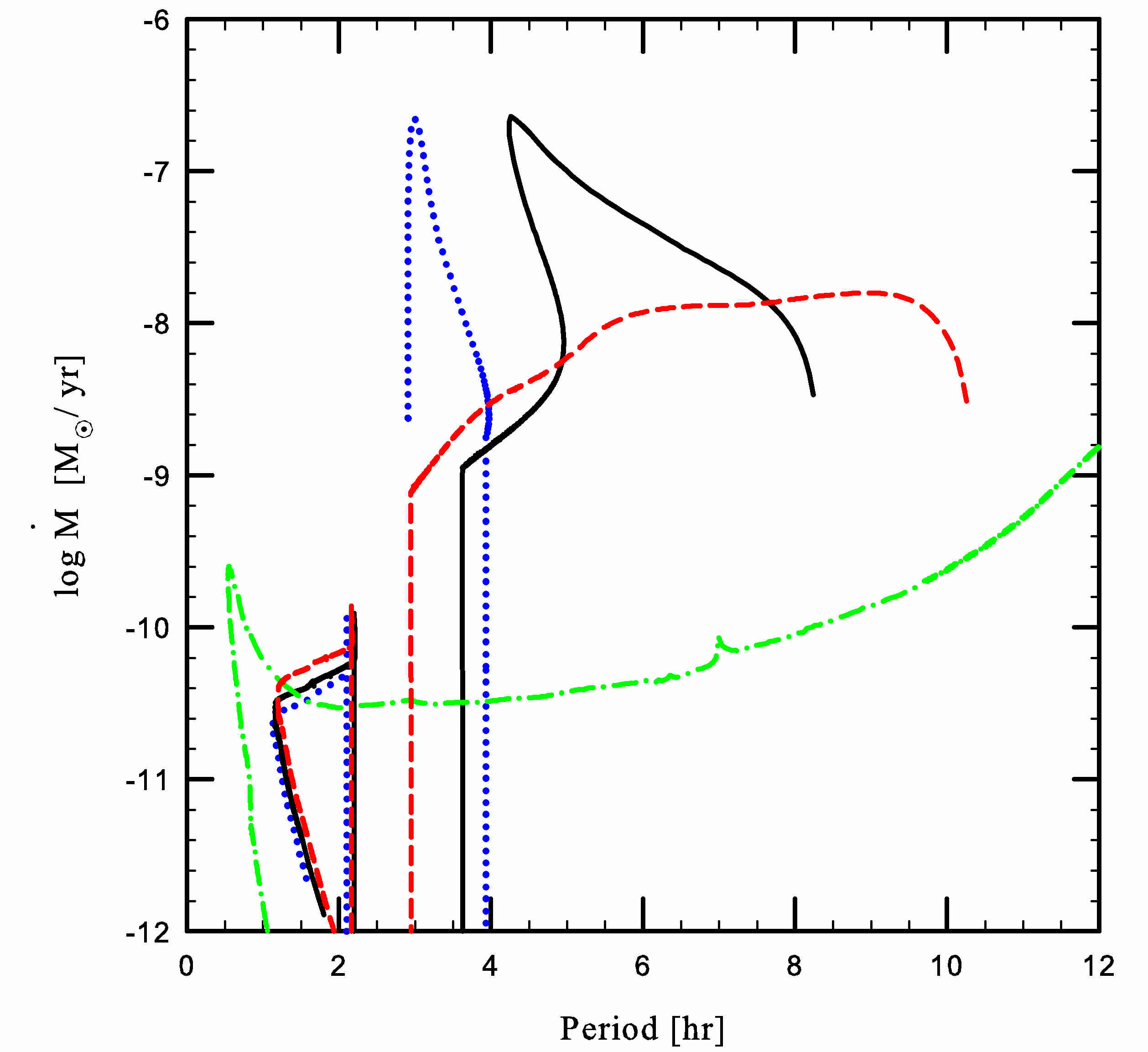}
\caption{Evolution of the mass-transfer rate ($\dot{M}$) with orbital period ($P_{orb}$) for representative evolutionary tracks. The initial values of $M_{WD}$ (in $M_\odot$), $M_{donor}$ (in $M_\odot$), and $X_{c0}$, respectively, are: 
$ 0.40, 0.40, 0.7$ (dotted [blue] curve);  
$ 0.60, 1.20, 0.7$ (solid [black] curve); 
$ 1.00, 1.40, 0.7$ (dashed [red] curve); 
$ 1.00, 1.80, 0.1$ (dash-dot [green] curve).
The dashed track illustrates the `canonical' CV evolution curve with a period gap from 2.8 hr to 2.2 hr. The solid line illustrates the evolution of a CV that is marginally stable with mass-transfer rates reaching $\approx$$10^{-6.5} M_\odot$/yr. In contrast, the track of a hydrogen depleted system (dash-dot line) exhibits significantly lower mass-transfer rates and has no discernable period gap.
}
\label{sample_tracks}
\end{figure}

Because of the importance of the evolution of $\dot M$ with respect to $P_{orb}$, we have plotted four illustrative tracks corresponding to four very distinct types of evolutionary behaviors that CVs can exhibit (see Figure 1): 

\noindent
1) A generic CV evolution curve corresponding to a 1.4 $M_\odot$ donor losing mass to a 1.0 $M_\odot$ WD is illustrated by the dashed line.  At $P_{orb} \simeq 2.8$ hr, the donor becomes completely convective and magnetic braking is switched off (IMB). Because the donor is bloated relative to its thermal-equilibrium radius, the decrease in AML (GR only) causes the cessation of mass transfer and the binary becomes detached. This corresponds to the upper limit of the period gap.  Mass transfer recommences at the lower limit of the gap ($P_{orb} \simeq 2.2$ hr). The donor eventually becomes so electron degenerate and has been driven so far out of thermal equilibrium that it expands as a result of continued mass loss. During this process the orbital period attains a minimum value ($P_{min} $) and subsequently evolves back to higher values of orbital period.

\noindent
2) The solid line illustrates the evolution of a CV that is marginally stable. The initial mass of the donor is 1.2 $M_\odot$ and it is losing mass to a 0.6 $M_\odot$ WD. The mass-transfer rate is driven to extremely high values ($\approx 10 ^{-6.5} M_\odot$/yr) and the upper limit of the period gap is increased to $\simeq 3.6$ hours. 

\noindent
3) The dash-dot line corresponds to the evolution of an evolved donor at the onset of mass transfer. The initial mass of the donor was taken to be 1.8 $M_\odot$ but its central hydrogen composition, $X_c$, had been reduced to 0.1.  Because the initial conditions for the system place it close to the bifurcation limit, the binary can evolve to an ultra-short orbital period of about 30 minutes. The binary also continuously transfers mass as it evolves through the `orbital period gap'.  This is typical of chemically evolved donors, regardless of the effectiveness of magnetic braking. Those that exhibit a period gap always enter the detached phase at low orbital periods (i.e., below the gap). The orbital period width of the inherent gap is greatly reduced (and in some cases negligible). This type of behavior has been noted in previous papers (see, e.g., Nelson \& Rappaport 2003).

\noindent
4) The dotted line illustrates the evolution of a binary containing a helium white dwarf (HeWD) accretor (0.4 $M_\odot$). The donor is sufficiently massive that the system is very close to being dynamically unstable. The initial $\dot M$ is in excess of $10 ^{-7} M_\odot$/yr and this drives the donor significantly out of thermal equilibrium\footnote{An analysis of thermal timescale mass transfer in CVs is given by, for example, Schenker and King (2002)}. This causes the period gap to start at approximately 4 hours. This type of evolution can have a non-negligible effect on the observed width of the gap. Also, because of the lower mass of the WD, the mass-transfer rate below the gap is diminished thereby decreasing the value of $P_{min}$ (see, Knigge et al. 2011, and references therein).

\subsection{PDCV 2D Probability Densities}

The probability densities of finding a PDCV with a particular pair of observables at the current epoch are shown in Figures 2 - 6. The vertical color bar on the right-hand side of the plots denotes the probability density on a logarithmic scale. Specifically, the probability density is a two-dimensional measure of the probability per unit interval in both the abscissa and ordinate variables of finding a PDCV with properties centered on those two intervals.  Since all of the intervals (pixels) have the same size, the color of a specific pixel is a relative measure of the probability of finding members of the ensemble with the properties defined by that pixel (the normalization of the color bar is arbitrary).


\begin{figure*}[! h]
\centering
\includegraphics[scale=1.55]{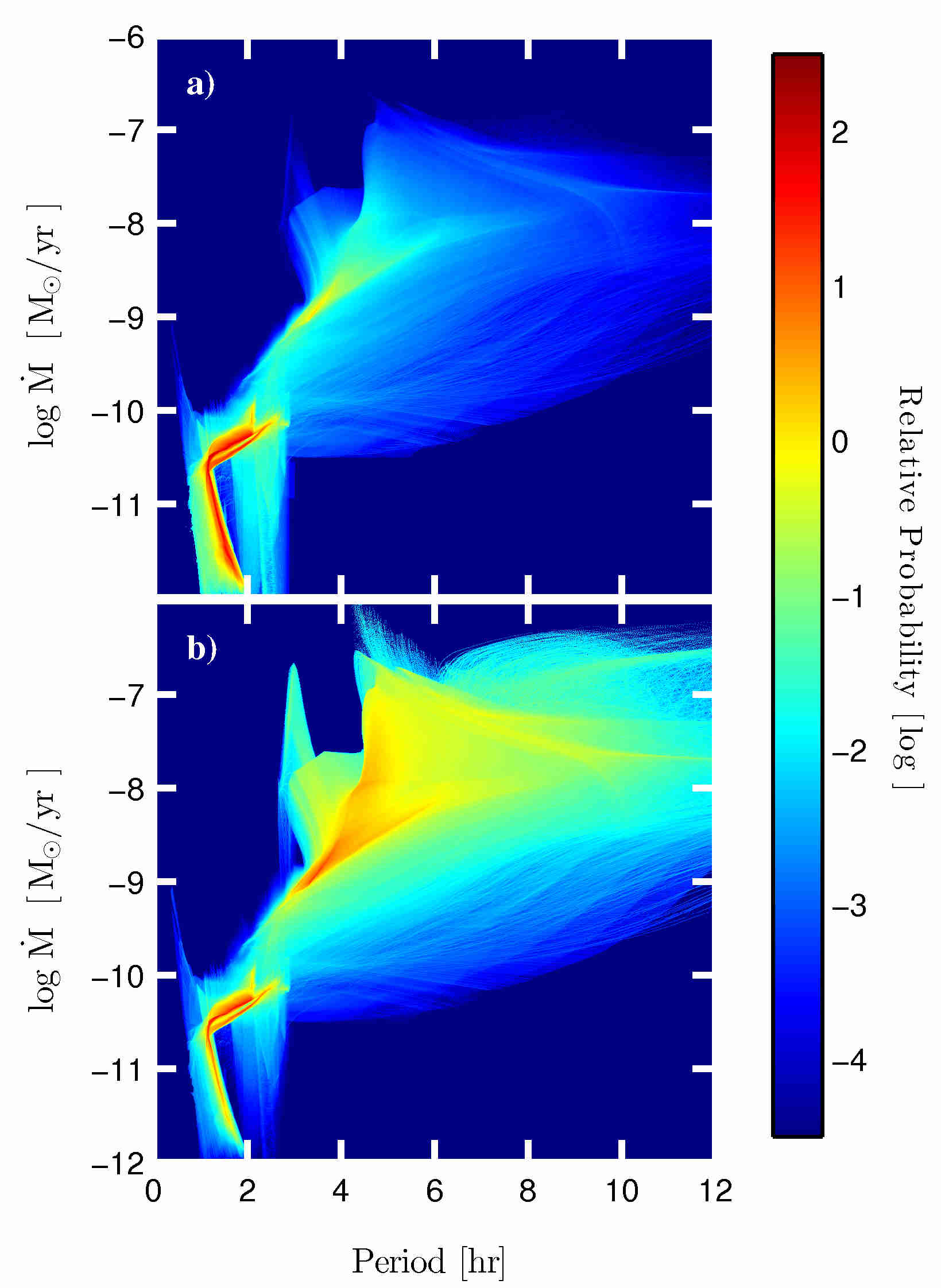}  
	
\caption{PS for PDCVs corresponding to our standard case (Case 1, see Table 1) in the $P_{orb} - \dot M$ plane. $\dot M$ is the mass-transfer rate and $P_{orb}$ is the orbital period of the system. The probability
density for a given combination of the two observables has been arbitrarily normalized and the vertical color bar on the right-hand side illustrates that probability on a logarithmic scale. Hence the color of a particular $ P_{orb} - \dot M $ pixel represents a relative measure of the probability that a PDCV has the properties defined by those two variables. Panels a) and b) each comprise 1000 horizontal cells covering an orbital period range of 12 hours and 1000 vertical cells corresponding to six orders of magnitude in the mass-transfer rate. The upper panel denotes the {\it intrinsic population} while the lower panel has been scaled by $\dot M$ to approximately take into account observational selection effects ({\it selected population}). 
}
\label{Mdot_Pcolor}
\end{figure*}

\subsubsection{Selection Effects}

Observational selection effects are an extremely thorny problem when it comes to understanding the ensemble properties of CVs. The underlying problem is associated with the fact that CVs can be grouped into many different subclasses most of which are discovered using a variety of different techniques. For example, some are discovered as a result of their brightness due to high mass-transfer rates while others are discovered due to the variation in the properties of their disks. In order to crudely examine possible observational selection effects, we have naively assumed that they are primarily due to a flux-limited optical passband that scales roughly as $\dot M^{2/3}$ (see HNR for more details). This in turn implies a flux-limited detectability that is proportional to $\dot M$.  Consequently we adopted a weighting scheme proportional to the mass-transfer rate.  We refer to the raw PS data as the {\it intrinsic population} and to the weighted PS results as the {\it selected population}. The bottom panels of Figures 2 - 6 show the results for the selected population. The most significant effect is a strong enhancement in the probability of detecting CVs with orbital periods above 3 hours because of the correspondingly high mass-transfer rates. This increase gives much more statistical weight to CVs found above the period gap (see the CV Catalog [V7.18] of Ritter \& Kolb 2003).

\subsubsection{$P_{orb} - \dot M$ Plane}

Figure 2 shows the results of the PS for PDCVs in the $P_{orb} - \dot M$ plane.  The superposition of the different behaviors described above can be clearly identified. To further understand what differences the inclusion of somewhat evolved CVs makes to the overall population, we have divided the results into two subsets (see Figure 3). The first subset includes CVs whose donors have not reached an age of 50\% of their respective terminal age main sequence (TAMS) ages at the onset of mass transfer (i.e., the unevolved donors in panel a)). The second subset includes only donors whose ages exceeded the 50\% delineator (i.e., the somewhat evolved donors in panel c)).  Figure 3 a) exhibits many of the features that have already been noted in previous PS studies (see, de Kool 1992; Kolb 1993; Howell et al. 1997; HNR). For example, we see a significant drop in the number of CV systems that should be observed with orbital periods between 2 and 3 hours (the period gap).  Also, CVs with periods above the upper limit have much higher mass-transfer rates than those below the period gap.  Once these systems enter a detached state, they cannot transfer mass and are no longer classified as CVs. However, systems can be born inside the period gap and they contribute to the number of systems that can be seen in the gap. Below the gap, the mass-transfer rate is significantly lower and monotonically decreases. The orbital period minimum for unevolved donors is located between $\approx 68$ to 74 minutes (depending on $M_{WD}$).  As has been noted by others, the vast majority of PDCVs located below the period gap follow two distinctly different tracks.  The upper track in Figure 2 (or 3 a,b)) consists of the thicker red line and corresponds to CVs with carbon-oxygen white dwarf (COWD) accretors while the lower one, that nearly parallels it, corresponds to CVs with HeWD accretors. The higher-mass COWDs enhance the rate of GR dissipation directly leading to higher mass-transfer rates and longer minimum periods.  For our standard case, no systems that have evolved past their minimum orbital periods (i.e., the period bouncers) ever reach an orbital period exceeding 2 hours because there is not enough time for them to evolve back into the gap.  If some form of supplemental AML were to be acting, $P_{min}$ could be increased substantially and CVs would have enough time to evolve back into the gap.  This possibility has been suggested by Knigge et al. (2011). 

Figures 3 c) and d) highlight the effects of the inclusion of nuclear evolution on the population of PDCVs. Some of the major qualitative differences that become apparent are: (i) the wider range of mass-transfer rates that are possible above the period gap; (ii) the much longer periods that can be attained ($> 10$ hr); (iii) the lack of a period gap for many of the evolved systems; and, (iv) the existence of very short-period CVs that can have high rates of mass transfer.  A more quantitative analysis for which the data has been folded into an orbital-period frequency distribution can be found in \S 3.2 and a discussion of the implications in \S 4. 


\begin{figure*}
\centering
\includegraphics[scale=1.00]{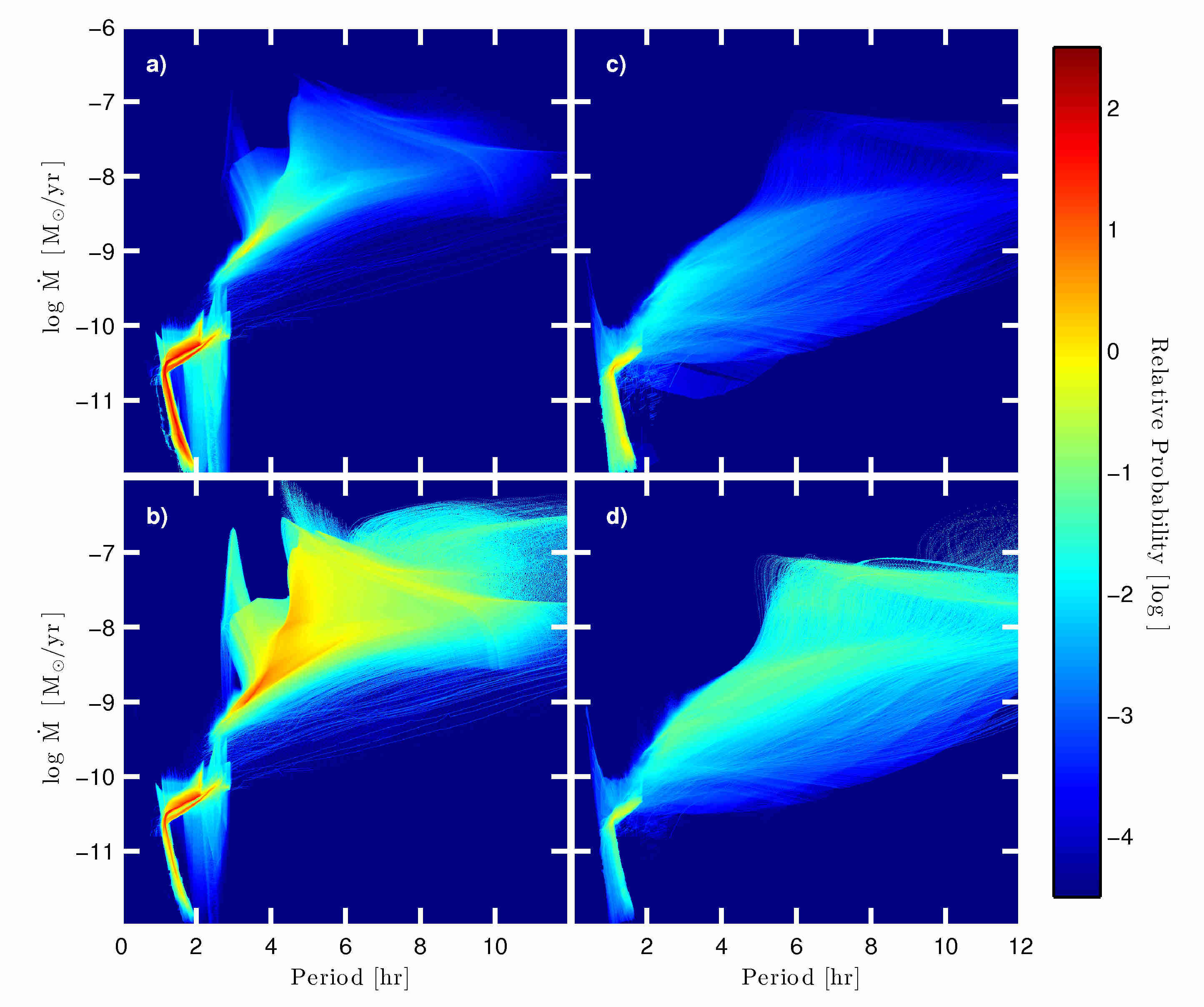} 
\caption{PS for PDCVs corresponding to our standard case (Case 1, see Table 1) in the $P_{orb} - \dot M$ plane. 
The population is divided into two distinct subsets and shown in separate panels. The unevolved group is shown in panels a) and b) and is composed of relatively young donors ($<$ 50\% of their respective TAMS ages) while the evolved group is shown in panels c) and d) and is composed of systems in which the donors have already undergone significant chemical evolution ($\geq$ 50\% of their respective TAMS ages). Combined the two groups make up the full PDCV population shown in Figure 2. 
The probability density for a given combination of the two observables has been arbitrarily normalized and the vertical color bar on the right-hand side illustrates that probability on a logarithmic scale. The upper panels, a) and c), correspond to the intrinsic population while the lower panels, b) and d), show the respective probabilities for the selected population. 
}
\label{Mdot_P_evol}
\end{figure*}

\begin{figure*}
\centering
\includegraphics[scale=1.00]{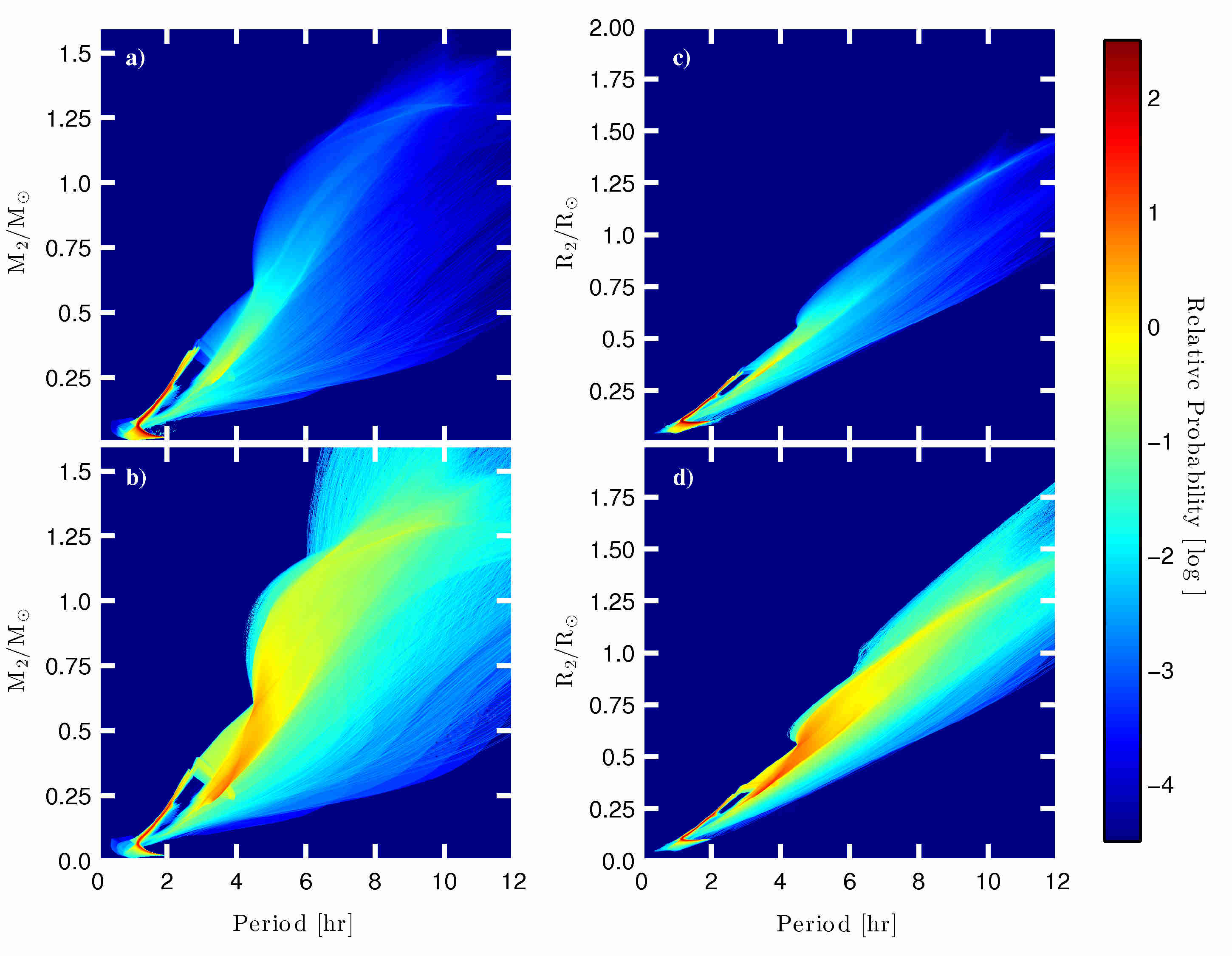}
\caption{PS for PDCVs corresponding to our standard case (Case 1, see Table 1) in the $P_{orb} - M_{2}$ plane (panels a) and b)) and in the  $P_{orb} - R_{2}$ plane (panels c) and d)). $M_{2}$ and $R_{2}$ are the mass and radius of the donor, and $P_{orb}$ is the orbital period of the system. The probability densities for a given combination of $P_{orb}$ and $M_{2}$ (or $P_{orb}$ and $R_{2}$) have been arbitrarily normalized and the vertical color bar on the right-hand side illustrates that probability on a logarithmic scale. Panels b) and d) show the respective probabilities for the selected population.
}
\label{M2_R2_color}
\end{figure*}

\begin{figure*}
\includegraphics[scale=1.00]{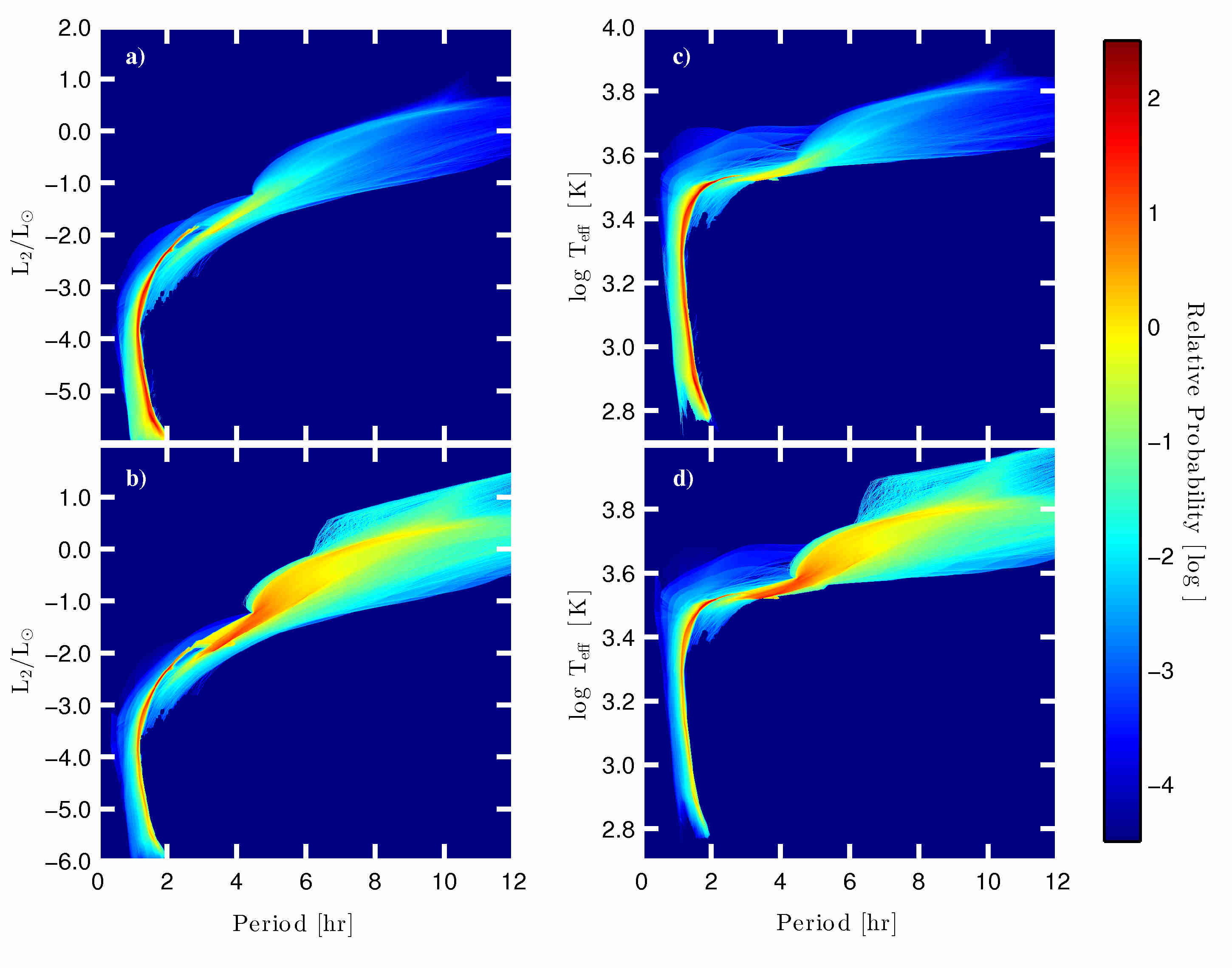}
\caption{PS for PDCVs corresponding to our standard case (Case 1, see Table 1) in the $P_{orb}$ - $L$ plane (panels a) and b)) and in the $P_{orb} - T_{eff}$ plane (panels c) and d)). $L$ and $T_{eff}$ are the luminosity and effective temperature of the donor, and $P_{orb}$ is the orbital period of the system. The probability densities for a given combination of $P_{orb}$ and $L$ (or $P_{orb}$ and $T_{eff}$) have been arbitrarily normalized and the vertical color bar on the right-hand side illustrates that probability on a logarithmic scale. Panels b) and d) show the respective probabilities for the selected population.
}
\label{L_T_Pcolor}
\end{figure*}

\begin{figure*} 
\centering
\includegraphics[scale=1.00]{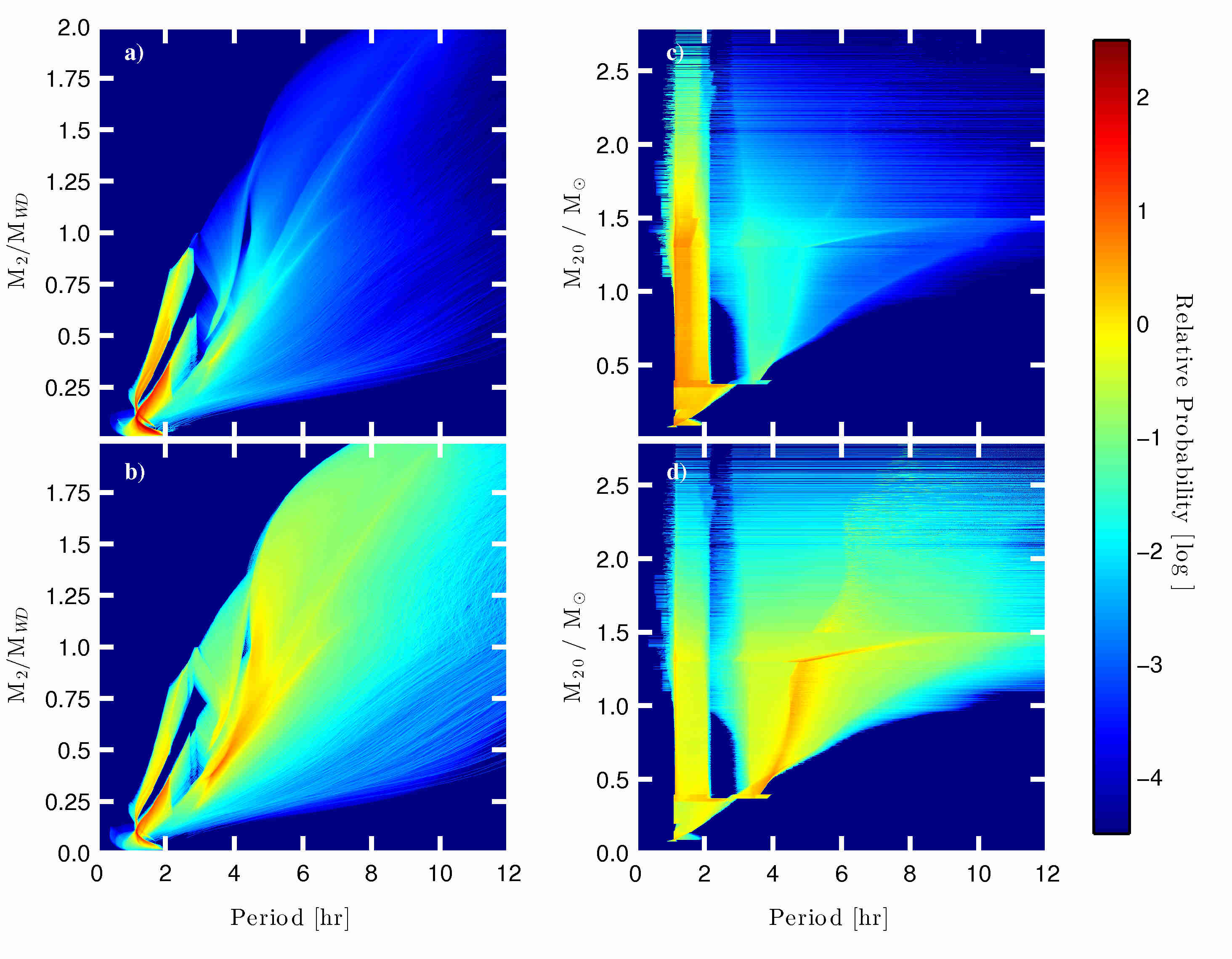}
\caption{PS for PDCVs corresponding to our standard case (Case 1, see Table 1) in the $P_{orb}$ - $M_{2}/M_{WD}$ plane (panels a) and b)) and in the $P_{orb} - M_{2 \mathrm{0}}$ plane (panels c) and d)). $M_{2}/M_{WD}$ is the mass ratio $q$ of the donor mass divided by the accretor mass, $M_{2 \mathrm{0} }$ is the initial mass of the donor when the system first comes into contact (ZACV), and $P_{orb}$ is the orbital period of the system. The probability densities for a given combination of $P_{orb}$ and $M_{2}/M_{WD}$ (or $P_{orb}$ and $M_{2 \mathrm{0}}$) have been arbitrarily normalized and the vertical color bar on the right-hand side illustrates that probability on a logarithmic scale. Panels b) and d) show the respective probabilities for the selected population.
}
\label{q_Mo_Pcolor}
\end{figure*}

\subsubsection{Other Observable Planes}

Figures 4 a) and b) show the probability densities for the mass of the donor as a function of $P_{orb}$, while Figures 4 c) and d) show the corresponding probability densities for the radius of the donor. Figure 4 a) implies that there are relatively few PDCVs whose donors have masses that are $\ge 1.0\, M_{\odot}$. We can also see two very distinctive triangle-like features in the lower-left corner of panels a) and b) for $P_{orb} \approx 1$hr. The upper feature (masses of approximately $0.09\, M_{\odot}$) is due to very low-mass donors that have just come into contact at the present epoch. These donors are extremely small (in radius) because they are near the end of the main sequence, and their orbital periods actually increase while they are losing mass. The other feature which is associated with donor masses about one-half as large is due to systems that contain highly-evolved donors.  They tend to evolve to much shorter minimum orbital periods before evolving back to longer periods. 

Figures 5 a) and b) show the probability densities for PDCVs in the (donor) luminosity-$P_{orb}$ plane. Figures 5 c) and d) are identical to the previous two (respectively) except that they show the effective temperature of the donor. It should also be noted that Figure 5 a) and b) differ significantly from Figure 7 (right panel) of HNR for $P_{orb} \geq 5$ hr. HNR found that the `probabilities' converged towards a single line for long-period systems whereas we find a wide range of luminosities for a given $P_{orb}$. This reflects the fact that some of our donors are ZAMS stars at the onset of mass loss while other donors are significantly evolved. With respect to the effective temperatures, we do see an analogous difference at long orbital periods. Our donors can have a wide range of effective temperatures largely because of nuclear evolution. These results are in excellent agreement with the spectral types derived by Beuermann et al. (1998).  The agreement is also very good for lower orbital periods where it is seen that the spectral types are later than would be expected for MS donors.

Figures 6 a) and b) illustrate the probability densities for the mass ratio $q$ as a function of orbital period. In agreement with previous PS studies, there are two distinct sets of tracks ($P_{orb} \leq 4$ hr) corresponding to higher and lower values of the mass ratio. Since the orbital period can be correlated reasonably well with the donor mass, and given that $q=M_{donor}/M_{WD}$, we can immediately deduce that the two different groupings correspond to two different sets of values for $M_{WD}$. The upper track corresponds to HeWD accretors and the lower track corresponds to systems containing COWD accretors.  Above the period gap we see a distinct border delineating a region beyond which no PDCVs exist.  The boundary separating these two regions has a value of $q \simeq 2.0$ at a period of about 8 hours and a value of $q \simeq 1.0$ at an orbital period of about 4 hr. The boundary occurs because of dynamical instabilities (see Nelson and Goliasch 2015). 

Figures 6 c) and d) show how the original mass of the secondary ($M_{20}$) correlates with the currently observed $P_{orb}$. Figure 6 c) is extremely instructive since it reveals several important aspects concerning the properties of PDCVs. For example, the red/orange vertical bar that can be seen for orbital periods between $\simeq$ 70 to 100 minutes confirms that most of the period bouncers can only evolve back up to an orbital period of about 100 minutes before they reach the current epoch (Case 1). This figure also shows that the contribution of primordial intermediate-mass donors ($>2.2\, M_{\odot}$) to the current population (below the bifurcation limit) is relatively unimportant.  It is also interesting to note that most PDCVs within the gap itself either started with low-mass donors or they had initial masses in the range of $\approx 1.0$ to $ \gtrsim 1.5\, M_{\odot}$. This latter result is not too surprising since it is the chemically evolved CVs that pass through the gap and their donors must be at least this massive to have undergone nuclear evolution.

The above analyses have also been applied to all of the other PS cases that we considered (e.g., different IMFs, BRFs). We do not show the color density plots for these cases because the {\it general} features described above for our standard case are repeated for these cases. While they share common features, the relative probabilities (and absolute numbers) can exhibit significant differences.  The implications for these cases will be analyzed (\S 3.3) by folding the data in the probability density plots into frequency distributions.  Also note that we have divided the data presented in Figures 4 - 6 into two subsets as was done for Figure 3 (unevolved and evolved cases).  These additional instructive figures and all high-resolution versions of those presented in this paper can be downloaded at http://www.star.ubishops.ca/CVsyn. 

\subsection{Orbital Period Frequency Distributions}

The frequency histogram (i.e., cumulative weight $W_{m}$) for all PDCVs as a function of $P_{orb}$ is shown in Figure 7.  As was the case for the color density plots, the 0 - 12 hour $P_{orb}$ range is divided into 1000 bins.  The upper panel (Figure 7 a)) gives the logarithm of the bin count of PDCVs while the lower panel (Figure 7 b)) has been weighted to depict the selected population (i.e., they are weighted with respect to $\dot {M}$).  For both panels, the diagonally-hatched histogram bins represent the entire PDCV population corresponding to our standard model (Case 1).  The cross-hatched histogram bins correspond to the subset of PDCVs for which the donors have undergone significant chemical evolution before the onset of mass transfer. Specifically, these systems have reached an age exceeding 50\% of their respective TAMS ages. The gray-shaded histogram bins correspond to the subset of PDCVs for which the donors have reached an age exceeding 80\% of their respective TAMS ages.  This group contains the most chemically evolved donors below the bifurcation limit. 

The period gap can clearly be seen in both panels.  The upper panel shows that there is more than one order of magnitude of PDCVs below the gap than above it.  However, the selected population (lower panel) indicates that more CVs would be observed above the gap.  We also see that there is a non-negligible percentage of CVs in the gap.  The lowest count in the gap at $P_{orb} \simeq 2.7$ hr has roughly the same count as is predicted for the intrinsic population at $P_{orb} \simeq 5$ hr.  Aside from the increasingly smaller fraction of highly evolved systems, Figure 7 also shows that the center of the period gap is shifted to lower values of $P_{orb}$ and becomes more narrow as the donors become more evolved (see, e.g., Nelson and Rappaport 2003 [NR]). Our results imply that more than 50\% of the non-magnetic CVs with orbital periods larger than 6 hours will be at least partially evolved (50\% TAMS).   

Figure 7 indicates that there is a roughly exponential drop-off in the numbers of PDCVs with respect to the orbital period for values {\it greater} than 6 hours. For the standard case, the {\sl e}-folding period is about 1.8 hours.  Thus
\begin{eqnarray}
{{dN} \over {dP_{orb}}} \propto e^{-{0.55}(P_{orb}/\mbox{hr})}, \qquad  P_{orb} > 6 \ \mbox {hr} \ . 
\end{eqnarray}
\noindent 
For the selected population, the e-folding period is longer ($\approx 2.2$ hours) because of the higher mass-transfer rates associated with longer-period systems.  Another feature that can be seen in Figure 7 (especially 7 b)) is the accumulation of systems with orbital periods between $\approx $4.5 to 5 hours.  These systems typically evolve from higher-mass ($\gtrsim 1.3 M_{\odot}$) donors with radiative envelopes that shrink rapidly when initially losing mass (with the concomitant decrease in $P_{orb}$) and then expand on their KH timescales.  These systems tend to have very high $\dot M$s, and thus some could be considered as supersoft X-ray sources.  If they were excluded, the shape of the distribution between $\approx$ 4.5 to 5 hours would be much flatter.  Furthermore, Figure 7 reveals that as the orbital period increases above 6 hr, PDCVs are increasingly more likely to be comprised of evolved donors.  CVs with evolved donors (as defined previously) are not a significant fraction of systems over any orbital period range except for the ultracompact CVs ($P_{orb} < 1$ hr). 


\begin{figure}
\includegraphics[scale=.66]{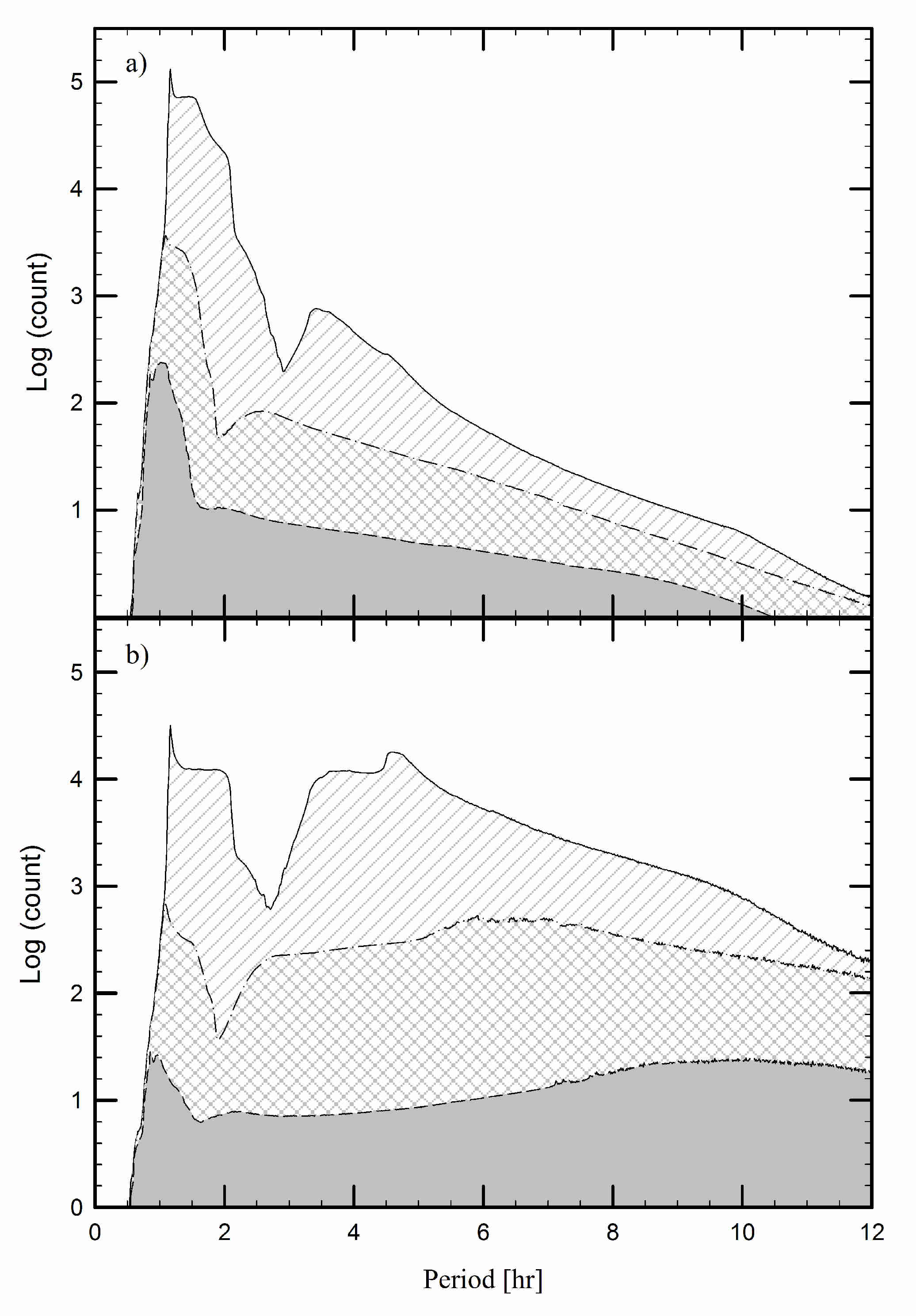}
\caption{Frequency histogram ($W_{m}$) for all PDCVs as a function of the orbital period corresponding to our standard case (Case 1, see Table 1). Panel a) shows the logarithm of the bin count of PDCVs while the lower panel (Figure 7 b)) has been weighted by $\dot M$ in order to approximately take into account observational selection effects. In both panels the diagonally-hatched bins contain all CV systems at the present epoch, while the cross-hatched bins and the gray-shaded bins only contain systems in which the donor star had an age of at least $50\%$ and 80$\%$ (respectively) of its terminal age MS lifetime when the system first came into contact. The bin width is 0.012 hr in both panels.
}
\label{Porb_binned_Case1}
\end{figure}

\begin{figure}
\includegraphics[scale=0.68]{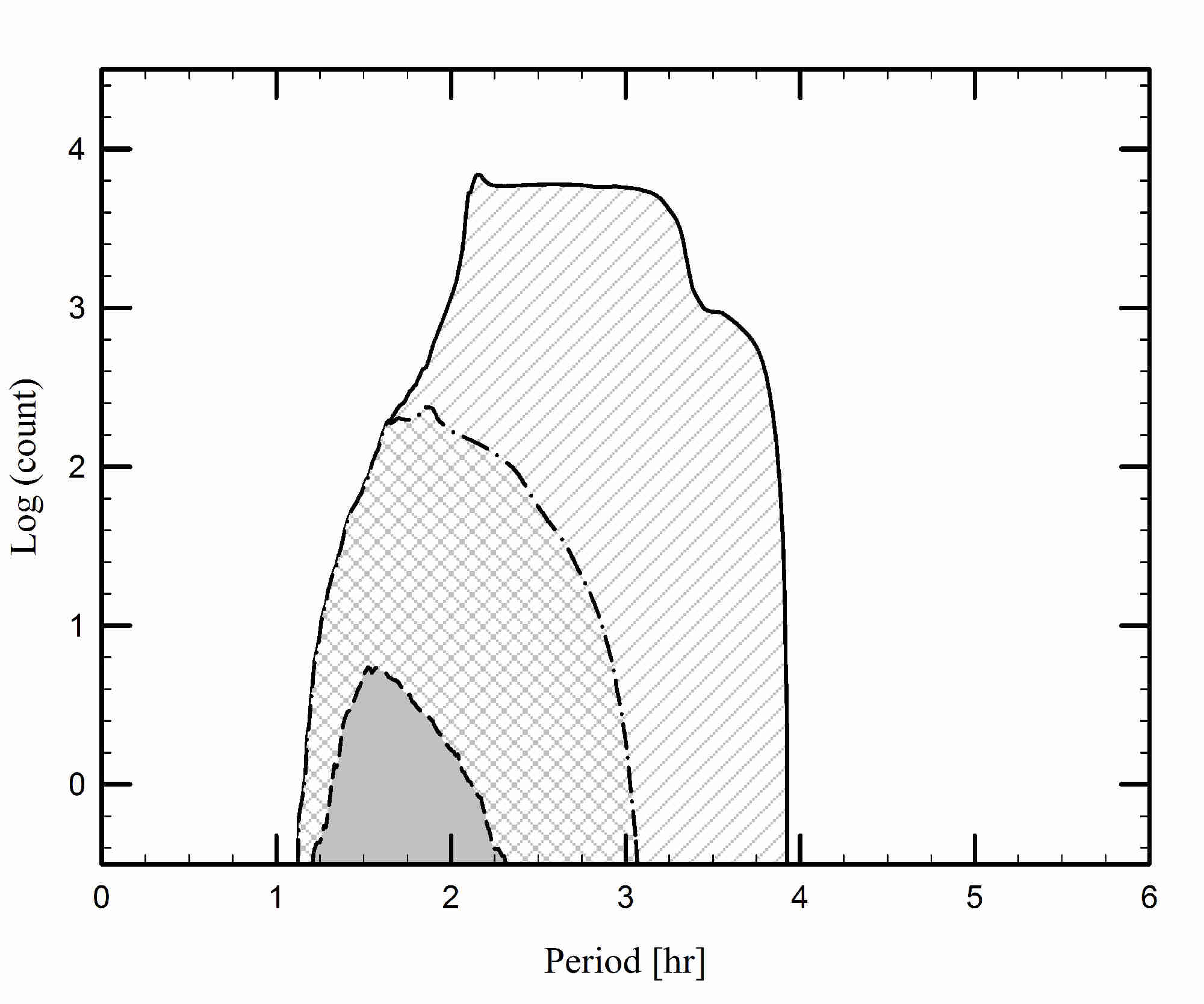}
\caption{Present Day (PD) orbital period distribution corresponding to our standard case (Case 1, see Table 1) of systems that are detached (in the period gap) and do not experience any mass transfer. The count in each bin represents the logarithm of the number of PDCVs to be expected in that particular period interval. The shading key for this figure is the same as for Figure 7. The bin width is 0.012 hr.
}
\label{Porb_Gap}
\end{figure}

\begin{figure}
\includegraphics[scale=0.68]{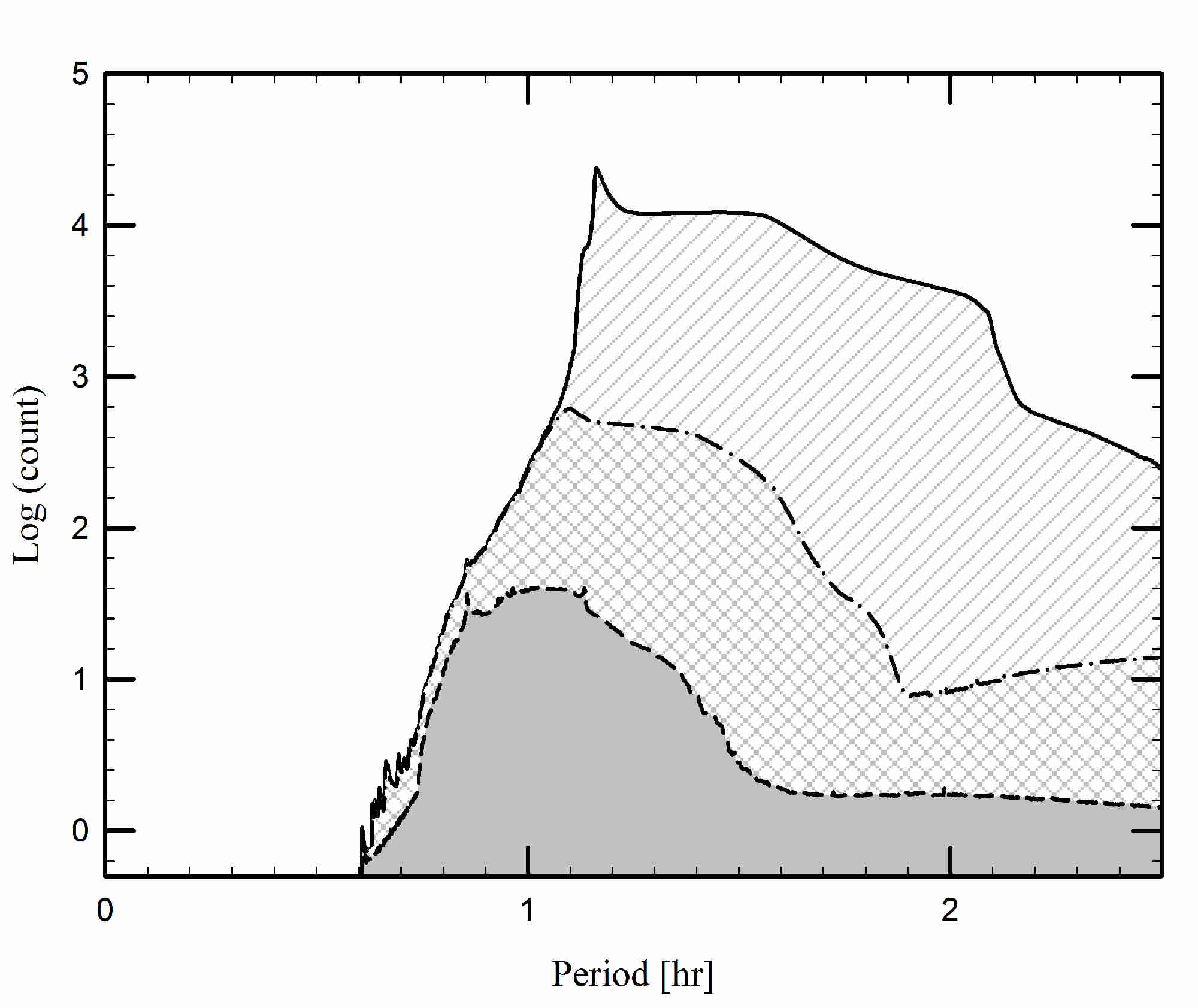}
\caption{PDCV orbital period distribution of short-period CVs corresponding to our standard case (Case 1, see Table 1). The details for this figure are the same as for Figure 7 except that the bin width is 0.002 hr.
}
\label{Porb_binned_ZOOM}
\end{figure}

In Figure 8 we show the numbers of evolved and unevolved present-day binaries that were once CVs and are now detached. These are systems that became detached as a result of IMB and are evolving across the gap as a consequence of AML due to GR. The figure strongly confirms the statements made above concerning the dependence of the location of the gap on the donor's composition. Note that this figure does not include the contribution from currently detached binaries that might become future ZACVs.     
Figure 9 is identical to Figure 7 a) except that the range of $P_{orb}$ is more suitable for the analysis of PDCVs located below the period gap (including the UCs).  The distribution exhibits a noticeable `spike' at a $P_{orb}$ of approximately 70 minutes, corresponding to the $P_{min}$ of generally unevolved donors with COWD accretors whose masses are primarily clustered around $0.6 M_{\odot}$. The spike is attenuated by the spectrum of WD masses and by the nuclear evolution of the donor.  The small lower period `bump' at $\simeq 68$ minutes corresponds to the $P_{min}$ of systems containing HeWD accretors.  The reason that a distinct spike is not seen is because the contribution from these systems is overwhelmed by that of CVs containing COWDs.  Below the gap we find that only $\simeq 19$\% of the PDCVs contain HeWDs.   Because the observational periods of CVs are typically only determined to within a precision of minutes, it would be extremely difficult to distinguish the two spikes that are predicted from our PS without a large unbiased data sample. This figure also makes it apparent that there is a reasonably sharp cut-off at the lower limit of the 
period gap. It is interesting to note that mildly evolved systems (50\% of TAMS) exhibit a maximum count at $P_{orb} \simeq 65$ minutes, considerably shorter than that found for largely unevolved donors with COWD accretors. The donors are somewhat hydrogen depleted and therefore have a comparatively smaller radii (implying shorter orbital periods). There are many fewer highly-
evolved donors but they are capable of reaching orbital periods of less than 30 minutes.  Although we cannot fully sample these systems because of their proximity to the bifurcation limit, it is clear that UC systems are much less likely to have formed. 

\subsection{Analysis of the Effects of Different PS Cases}

In Table 2 we examine the distribution and absolute numbers of PDCVs for some of the more interesting PS cases (both the intrinsic and selected populations that we have investigated).  The third and fourth columns give the absolute numbers of PDCVs and detached binaries (in the gap) that are predicted to be in the galactic disk, respectively.  The total number of PDCVs was calculated based on the estimated birth-rate of WDs in the Galaxy. Based on the analysis presented in \S 2.2.3, we assumed that a constant birthrate would produce $0.4\,\mbox{WDs}/\mbox{yr}$. According to the cases presented in Table 2, this yields a CV number density of $\approx 1 \pm 0.5 \times 10^{-5}$ pc$^{-3}$ at the current epoch.  This value is in reasonable agreement with the observations (Pretorius et al. 2007), but could easily be revised by a factor of two.  The next four columns enumerate the percentages of PDCVs that have orbital periods above the period gap, in the period gap, below the gap (but not yet beyond the minimum period), and past $P_{min}$ (i.e., the period bouncers). The last column indicates the percentage of HeWDs in PDCVs.


\begin{figure}[h!]
\centering
\includegraphics[scale=0.70]{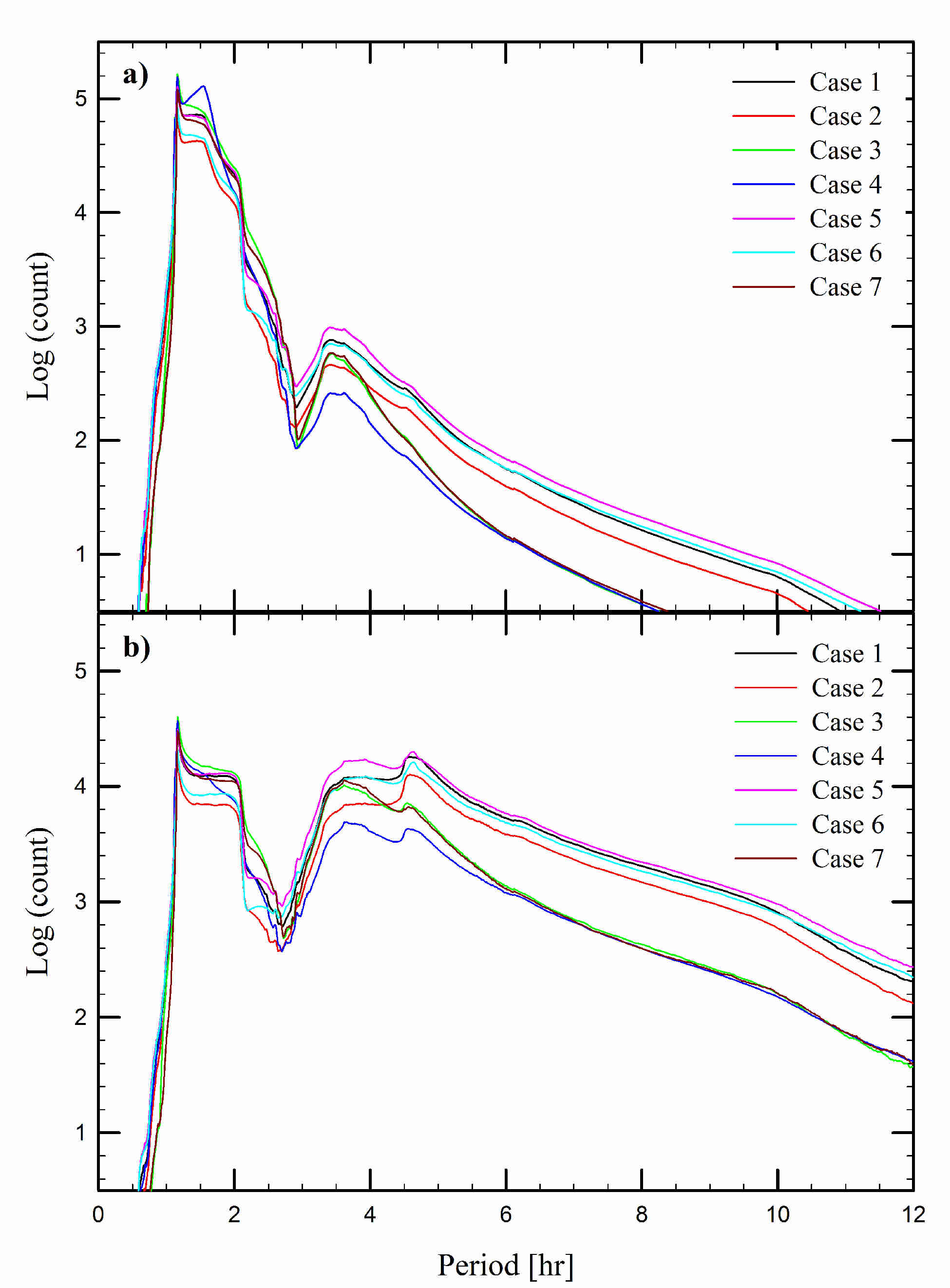}
\caption{PDCV orbital period distribution for various cases (differentiated by color). The Case numbers in the legend correspond to those in Table 1. Panel a) shows the logarithm of the bin count of PDCVs while the lower panel (Figure 7 b)) has been weighted by $\dot M$ in order to approximately take into account observational selection effects.  The bin width is 0.012 hr in both panels.
}
\label{Porb_Cases}
\end{figure}


\begin{deluxetable*}{ccccccccc}

\tablecaption{\label{tab:Period Distribution} Properties of PDCVs}

\tabletypesize{\scriptsize} \tablewidth{0pt}

\tablehead{ \colhead{Case}& \colhead{}& \colhead{Present-day}& \colhead{In Gap\tablenotemark{\dag}}&
\colhead{$P >$ 2.8 hr}& \colhead{$P \le$ 2.8 hr }& \colhead{$P \le$ 2.2 hr }& \colhead{ $P \ge$ $P_{min}$}& \colhead{Fraction} \\
\colhead{\#}& \colhead{}& \colhead{CVs\tablenotemark{\dag}
}& \colhead{(Detached)
}& \colhead{}& \colhead{$P \ge$2.2 hr}& \colhead{$P >P_{min}$}& \colhead{(Bouncers)}& \colhead{of HeWDs}  }
\startdata
& & & & & & & & \\
1& intrinsic& 4.54E+06& 6.56E+05& 2.0\% & 1.7\% & 54.1\% & 42.2\% &  18.9\% \\
 & selected& & & 74.3\% & 1.2\% & 21.3\% & 3.2\% & 10.4\% \\
& & & & & & & & \\
2& intrinsic& 2.60E+06& 3.99E+05& 2.2\% & 1.4\% & 52.8\% & 43.6\% & 15.6\% \\
& selected& & & 77.6\% & 1.0\% & 18.5\% & 2.9\% & 8.8\% \\
& & & & & & & & \\
3& intrinsic& 5.45E+06& 4.81E+05& 0.9\% & 2.8\% & 57.8\% & 38.5\% & 30.3\% \\
& selected& & & 50.7\% & 3.2\% & 40.6\% & 5.4\% & 21.5\% \\
& & & & & & & & \\
4& intrinsic& 6.12E+06& 3.07E+05& 0.5\% & 1.2\% & 38.5\% & 59.9\% & 24.6\% \\
& selected& & & 45.2\% & 2.3\% & 41.4\% & 11.2\% & 29.5\% \\
& & & & & & & & \\
5& intrinsic& 4.53E+06& 7.93E+05& 2.5\% & 1.8\% & 54.0\% & 41.7\% & 13.0\% \\
& selected& & & 76.7\% & 1.3\% & 19.3\% & 2.8\% & 12.1\% \\
& & & & & & & & \\
6& intrinsic& 3.04E+06& 5.68E+05& 2.9\% & 1.5\% & 53.8\% & 41.8\% & 13.1\% \\
& selected& & & 79.4\% & 1.1\% & 17.0\% & 2.5\% & 12.5\% \\
& & & & & & & & \\
7& intrinsic& 4.17E+06& 4.73E+05& 1.2\% & 3.1\% & 54.7\% & 41.0\% & 13.4\% \\
& selected& & & 57.1\% & 3.2\% & 34.8\% & 4.9\% & 15.9\% \\
& & & & & & & & \\

\enddata

\tablenotetext{}{\textsuperscript{\dag}Absolute number in the disk normalized to a constant BRF of 0.4 WDs/yr.}

\end{deluxetable*}

As expected, the fraction of systems observed above the period gap is significantly higher for the selected populations.  We see that only 2\% of Case 1 PDCVs are found above the gap, 1.7\% evolve through the gap, and the rest reside below the gap. For the selected population, it is about three times more likely to detect CVs above the period gap.  The results for Cases 2 to 7 are also tabulated. The fraction of HeWDs can vary over a span of about 13\% to 30\% (intrinsic population) and the fraction of PDCVs above the gap varies between 0.5\% to 3\%. Finally we note that the number of PDCVs that have evolved beyond the minimum period is always about  40\% of the total, regardless of the case. Although this is a large fraction, they are notoriously difficult to detect because of their faintness (not surprisingly they only constitute a few percent of our observationally selected population).

In Figure 10 the frequency distributions with respect to $P_{orb}$ for Cases 1 through 7 are contrasted. Panel a) applies to the intrinsic populations while panel b) corresponds to the selected populations. Not unexpectedly, the {\it general} shape of the $P_{orb}$ distributions is very similar for all cases, but the absolute number of CVs can vary significantly depending on the case. Other features, such as the minimum period spike, can either be accentuated or diminished depending upon the physical conditions inherent for each case. The most striking difference occurs for the exponential birthrate function (Case 5). Compared to our standard case (1), the probability of detecting CVs (intrinsic population) above the period gap is reduced by nearly a factor of 4. The reason for this difference can be attributed to the fact that for an exponentially declining BRF most CVs are very old, and therefore we would expect to find fewer CVs evolving through the relatively rapid phase of mass-transfer evolution for orbital periods above the gap.

The results for Case 3 are also interesting to analyze because of the much stronger correlation between the primordial primary and secondary masses.  We would expect to have many more massive (and evolved) donors at the onset of mass transfer.  Since massive and/or evolved donors tend to have larger radii this typically implies longer orbital periods (this is especially true for Case 6).  We would also expect to observe more UC CVs; although the fraction of these systems is very small ($<1\%$), we do see the enhanced contribution of the initially more massive (and therefore more likely to be evolved) donors at the onset of mass transfer. However, when the primordial component masses are uncorrelated, we see the opposite effect.  Less than 0.1\% of ZACVs are capable of evolving into UCs, and most of these will have orbital periods not much less than 60 minutes.

When $\epsilon=0.33$, we see relatively little difference in the distributions with respect to the standard case. This is especially true for the observationally selected population. The value of $\epsilon$ is poorly known and thus is a significant source of uncertainty. Our results show that the shape of the distribution seems reasonably robust with respect to this uncertainty. However, we also note that Politano and Weiler (2007) found that significant differences only occur for smaller values of $\epsilon$ (which are also physically possible). However, the CE formulation that Politano used was also somewhat different from the one adopted here, and thus a direct comparison is not easily accomplished.

Cases 8 and 9 are excluded from the above discussion as they did not yield a significant number of successful present-day CV systems. With respect to Case 8 we note that our result confirms the general trend for low values of common envelope efficiencies found by Politano and Weiler (2007).  A large decrease in common envelope efficiency significantly reduces the number of systems emerging from the CE phase as potential CVs.  For Case 10 we found that the number of PDCVs with evolved secondaries was reduced because of the relatively higher number of low-mass systems. This result is what we would expect using the Salpeter IMF because its power-law index grossly overestimates the formation of low-mass stars. Case 11 simulates a ``star-burst"� in which all stars are born contemporaneously with the formation of the Galactic disk.  This produces a radically different distribution of PDCVs for which almost all of the systems ($\sim 99\%$) have orbital periods of less than 2.2 hours.  This result was also expected because in such a scenario most CV systems have had a significant amount time to evolve and thus pass through their period gap.  A more attenuated version of a star burst is simulated by Case 4. It allows for a richer analysis of such a scenario and was illustrated above.

\section{\label{sec:DISCUSSION}DISCUSSION}

The results of our PS study are in broad agreement with those of HNR (and the sophisticated analysis by Podsiadlowski et al., 2003).  Most of the prominent features are reproduced, but this study is much richer than that of HNR in terms of the behaviors exhibited by CVs due to the inclusion of the nuclear evolution of the donor and our ability to account for high-mass secondaries (donors).  In particular, we see that the ranges of orbital periods, mass-transfer rates, and the properties of the donor cover a much wider range of values. 

Based on the observed fraction of Nova-likes (NLs) above the period gap, we would expect $\dot M$ to be $\gtrsim 10^{-8}\, M_{\odot}/\mbox{yr}$.  Figure 2 certainly supports this claim although there is a population of evolved systems that have much lower mass-transfer rates (see the dash-dot curve in Figure 1).  The evolved systems might thus account for the observed fraction of Dwarf Novae.  Figure 2 also shows that some systems can have mass-transfer rates in excess of $10^{-7}\, M_{\odot}/\mbox{yr}$ and that they typically have $P_{orb} > 4$ hr. These systems should probably be classified more aptly as supersoft X-ray sources. We do not attempt to distinguish them from other classes of PDCVs.  But we also see many CVs that are transferring mass at rates in excess of $10^{-8}\, M_{\odot}/\mbox{yr}$ that have orbital periods between approximately 3 and 4 hours.  For these systems, the donor star typically has a mass very similar to that of the WD and thus is on the verge of dynamical instability.  There is also a large number of CVs with $P_{orb} > 4$ hr that contain COWDs which have even higher mass-transfer rates.  Both of these types of systems are potentially very interesting because they may be related to the SW Sextantis sub-class of NLs (see, Hoard et al. 2010, and references therein). 

As has been noted in previous papers (e.g., Howell et al. 1997), HeWDs can comprise a significant fraction of the total number of WDs.  For the range of cases that we examined, the fraction of PDCVs containing HeWDs is between 13 and 30\%.  However, there is no observational evidence for the existence of HeWDs in CVs.  For almost all of the cases that we analyzed, the masses of the COWDs were peaked at $\approx 0.65 M_\odot$.  For most cases, the number of systems that contain WDs with masses in excess of $1.2 M_\odot$ is negligible.  But for certain combinations of the PS parameters $\epsilon$ and $\xi$, we find that the WDs in PDCVs can have masses close to the Chandrasekhar Limit.  For Cases 7 and 5, the fraction of high-mass WDs ($> 1.2 M_\odot$) can be as large as 0.5 and 1\%, respectively.  Thus we would expect to find a larger percentage of Recurrent Novae in these populations.

\subsection{Period Gap}

The IMB paradigm is currently the most accepted explanation for the period gap (see, e.g., Kolb et al. 1998), often quoted as being between 2.2 and 2.8 hours. The physical reasons for the ineffectiveness of MB when the donor loses its radiative core are not clear.  Even though low-mass (convective) stars normally exhibit magnetic phenomena (e.g., starspots), it is assumed that the topology of the B-field changes drastically and can no longer be efficient in transporting angular momentum out to the corotation radius.  While there have been other hypotheses put forward to explain the existence of the gap (Taam et al. 2003), the IMB is still very viable (Kolb et al., 1998).  

The synthesized gap for all of our PS cases is wider than that observed.  The main reason for this is that the populations contain a significant number of low-mass, but not yet convective, donors ($\ge 0.37 M_\odot$) that experience high mass-transfer rates when they initially overflow their lobes (Kelvin-Helmholtz mass transfer).  This drives the binaries to higher orbital periods before MB is switched off (see the dotted curve in Figure 1). The cumulative  effect is that the upper limit of the period gap is raised.  In retrospect, we could have easily narrowed the width of the synthesized gap (and hardly change $P_{orb}$ at the lower edge) by arbitrarily reducing the magnitude of the MB in the VZ law.  

Contrary to the results of HNR, we find that there are many more CVs in the gap (see their Figure 10). The smallest number of systems according to the PS occurs at a $P_{orb} \simeq 2.9$ hr.  Figure 7 shows that for some values of $P_{orb}$, (partially) evolved donors can account for nearly 30\% of the systems.  But the vast majority are CVs that were born within the gap.  This is especially true for populations where the secondary mass is uncorrelated with that of the primary (Cases 3 and 7).  These cases allowed for the inclusion of many lower-mass secondaries as dictated by the MS IMF.  As mentioned previously, it would be hard to distinguish these H-rich systems from evolved donors because their surface abundances of hydrogen will not be observationally distinguishable (except for extreme cases).  Finally, we note that magnetic CVs (polars) probably do not undergo IMB.  They would evolve through the gap and do constitute a substantial fraction of the CVs that are observed there.  

\subsection{Minimum Period ``Spike''}

Kolb \& Baraffe (1999) pointed out the fact that the existence of an orbital period minimum implies that there should be a `spike' in the number of systems whose orbital periods are coincident with that period ($\dot P_{orb} =0\Rightarrow {dN}/{dP_{orb}} \rightarrow \infty$ with the width of the spike becoming infinitesimal).  Although the empirical period distribution is fraught with selection effects, an increase in the number of systems located near the observed minimum period has been reported (see, e.g., G\"{a}nsicke et al. 2009) based on data obtained using SDSS observations.  If one assumes that all WDs have the same mass and that the donors are completely unevolved (ZAMS) stars, then the spike would have to be extremely pronounced\footnote{We have assumed that the only sink of angular momentum is due to GR. Thus the rate of AML is fixed by the masses of the donor and the WD, as well as the orbital period.}. However, the fact that WDs can have different masses and that the donors will have chemical compositions different than what would be expected for ZAMS stars, implies that this spike will be attenuated.  

For our standard case, there is a significant enhancement in the probability of discovering CVs near the minimum orbital period. Specifically, we find that the probability density (per unit $P_{orb}$) of detecting CVs at $P_{min}$ will be at most twice as large as the probability density for detecting CVs in the period range of $\approx$ 76 to 106 min (see Figure 9). For most of the cases that we investigated, we found that the width of the spike at its base was nearly 5 minutes (see, e.g., Figure 9) and that the {\it maximum} probability density was increased by nearly a factor of two relative to what would be detected for orbital periods extending up to 85 minutes.  We also examined how the minimum periods were affected by the mass of the WD accretor. We found that the minimum orbital periods deviated by approximately 1.8 minutes when comparing a $0.4 M_\odot$ HeWD accretor with a $0.6 M_\odot$ COWD accretor. If we take the mass of the COWD accretor to be as large as $1.0 M_\odot$, the value of the minimum orbital periods increases by another 2.2 minutes. Given the fact that most of our PDCVs contain COWDs with masses close to the canonical value of $0.6 M_\odot$, we conclude that the chemical evolution of the donor has also played a role in attenuating the minimum period spike. 

Another interesting feature that becomes apparent in Figure 2 is the very sharp drop in $\dot M$ just as systems evolve through their minimum orbital periods. As they evolve toward $P_{min}$, we see that, for most CVs, $d \log \dot M / d P_{orb} \simeq +0.28$/hr, while as they evolve away from $P_{min}$ we have $d \log \dot M / d P_{orb} \simeq -2.0$/hr.  The junction between these two regimes is not characterized by a discontinuous change in the slope.  Instead we see a region that is about 3 minutes wide (coincident with the $P_{min}$ spike) where it is very difficult to distinguish CVs on either side of $P_{min}$. The value of $\dot M$ decreases by almost a factor of 1.8 for orbital periods within $\approx 3$ minutes of the minimum orbital period.  If systems were selected largely based on their accretion luminosity, then we might reasonably expect the brightness to decrease by almost a factor of two over this orbital period range. This corresponds to a decrease in brightness of nearly 2/3 of a magnitude which may contribute to the difficulty in making unambiguous detections of period bouncers (see, e.g., Aviles et al. 2010).  Despite this problem, we are convinced that they must exist; otherwise the canonical model for the evolution of CVs requires serious revision.  This is an extremely important issue and needs to be addressed more fully by future surveys. 

\subsection{Ultracompact Systems}

Although the observationally determined orbital period distribution of CVs shows a reasonably well defined cutoff at approximately 80 minutes, nearly 30 UC systems have been discovered (including the AM CVn stars). This constitutes about 3\% of all binaries that could be broadly classified as CVs.  We define ultracompact CVs as those having orbital periods of less than 60 minutes and comprised of at least one WD.  UCs have been extensively investigated by Sienkiewicz (1984), Nelson et al. (1986), Podsiadlowski et al. (2002), Bildsten (2002), Nelemans et al. (2010), NR (and references therein).  Four different evolutionary channels could lead to the formation of such systems. These include: (i) a double degenerate channel that produces two WDs in a very tight orbit after one or two CE phases (a HeWD is typically the donor); (ii) a partially degenerate helium-star that is losing mass to a COWD accretor after several previous episodes of mass transfer; (iii) a brown dwarf for which mass transfer is initiated while it is degenerate; and, (iv) a channel that corresponds to the UC tracks in our evolutionary grid.  According to this latter scenario, binary systems that have initial conditions that place them close to the bifurcation limit will have donors that are already rich in helium in their cores.  They contain such a large fraction of helium that they are considerably smaller than comparable (unevolved) donors and thus can attain substantially lower minimum periods.

Based on computations that we previously carried out (NR and NDM), we found that it is possible for CVs to reach orbital periods as short as 6 minutes. This is sufficiently short to explain the orbital periods of all AM CVns discovered to date. We found that the systems with the shortest possible orbital periods will only have miniscule hydrogen surface abundances ($X_s \lesssim 10^{-3}$), and other than their somewhat thermally expanded radii, they are virtually indistinguishable from HeWD donors. The question is whether the `CV channel' can make any significant contribution to the observed number of ultracompacts. The problem with this particular channel is that the initial conditions must be {\it extremely} finely tuned in order to attain these short orbital periods. Unfortunately, the extremely limited volume of the phase space of initial conditions required to model these systems causes our grid to be too coarse to obtain a statistically meaningful sample of systems with $P_{orb} \lesssim 30$ minutes.  
Nonetheless, Figure 9 clearly shows that there is a very precipitous decline in the number of PDCVs (intrinsic population) with orbital periods of less than one hour. This decline is so steep that we can probably safely claim that the CV channel should make a negligible contribution to ultracompacts with $P_{orb} < 30$ minutes. In fact, the contribution to UCs for our standard case is even small for higher orbital periods but it should be noted that the fraction of UCs is quite sensitive to the value of $\xi$.  If $\xi =1$ (reasonably strong component correlation), then the absolute numbers of PDCVs drops (see Table 2) but the fraction of UCs can be as large as $\approx 1\%$. If the component masses are not correlated, the fraction of UCs decreases to considerably less than 0.1\%. Thus it is difficult to make any definitive quantitative statement, but we can say that the CV channel probably does not contribute significantly to the population of observed UCs (especially for $P_{orb} \lesssim 30$ minutes) unless selection effects are a dominant consideration.

\subsection{WD Mass Accretion}

The issue of whether or not WDs in CVs actually gain or lose mass during their evolution can be addressed by population syntheses. In order to carry out self-consistent calculations, the amount of mass that is either accreted (or eroded) must be incorporated into the computation of the CV's evolution. According to the grid of accreting WD models generated by Prialnik \& Kovetz (1995) and more recently by Yaron et al. (2005), the amount of mass gain or erosion depends primarily on $\dot M$ and to a lesser degree on the mass and internal temperature of the WD (see also Nelson et al. 2004b).  It is well known that if $\dot M$ is of the order of $10^{-6}$ to $10^{-7} M_\odot$/yr then quasi-steady nuclear burning can occur, as is observed in supersoft X-ray sources (see, e.g., Van den Heuvel et al. 1992).  The net result is that this allows COWDs to grow in mass. The magnitude of this increase and its sustainability clearly have major implications for the validity of the so-called single degenerate scenario for Type Ia supernovae. As $\dot M$ decreases, the amount of accreted material that remains on the surface of the white dwarf diminishes. For mass-transfer rates of less than $\sim 10^{-10} M_\odot$/yr, significant erosion starts to occur. 

Ritter \& Burkert (1986) concluded that the measured high mean masses of WDs in CVs (and the absence of HeWDs) could be explained as a selection effect. A very careful study of this issue was recently carried out by Zorotovic et al. (2011) and they conclude that the masses of COWDs in PDCVs actually increase by $\lesssim 0.2 M_\odot$ (assuming that the natal masses of WDs in CVs are no different from those observed in the field). Thus they conclude that WDs are not eroded in mass (on average), but rather gain mass. 

One way to attempt to address this question would be to recalculate our grid of evolutionary models taking into account the expected mass gain or loss of the WD while self-consistently redistributing the orbital angular momentum. Our PS could then be re-computed and an inference could be obtained. In order to obtain a rough estimate of $\delta M_{WD}$, we interpolated the models of Yaron et al. (2005) to infer the value of $\beta$ as a function of $M_{WD}$ and $\dot M$. We note that the value of $\beta$ is uncertain. Although we could not alter the tracks in our grid, we know that the {\it general behavior} of the tracks is not greatly altered depending on whether the mass transfer is partially conservative or partly erosive (assuming that $| \beta |$ is not close to unity).  Based on our grid ($ \beta  = 0$), we find that $\langle \delta M_{WD} \rangle$ falls in the range of $\approx 0.05-0.10 M_\odot$ (positive mass gain of PDCV WDs). This range was determined for our standard case but we allowed the WD to have temperatures of between 10 and 50 million Kelvin.  The other cases were also analyzed and for each one there was a positive mass gain.  
 
As a cautionary check of our analysis, we have re-computed a few of the `canonical' CV tracks that self-consistently included the effects of mass erosion and mass gain on the evolution.  We conclude that our estimate of $\delta M_{WD}$ would have to be somewhat lower.  On the other hand, if we accept the results of Yaron et al. (which admittedly have their own uncertainties), we believe that the mass gain could easily be larger because our grid did not fully sample tracks at high mass-transfer rates.  It is the relatively-short, initial mass-transfer phase that can lead to very high rates of sustained mass transfer (especially for higher-mass donors) during which time a significant fraction of the mass of the donor is actually accreted by the WD. It is worth noting that below the period gap, where supposedly significant erosion will occur, the mass of the donor is so small that even though it might lose between $0.12 - 0.15 M_\odot$ of its mass, the WD can only lose a similar amount (or less) of its own mass.  Thus our very preliminary analysis is highly suggestive that WDs in PDCVs may have gained as much as $0.1 M_\odot$.  We plan to address this issue more fully in a future paper.

\section{CONCLUSIONS}

Using a highly efficient approach to population synthesis, we pre-compute representative tracks for the evolution of CVs and then interpolate the grid for a specific set of initial conditions corresponding to the properties of a particular ZACV. A similar approach to population synthesis has been used by Podsiadlowski et al. (2003) and by Willems et al. (2005, 2007); however, both used  stellar models whose input physics was not as intricate. Since the interpolations are relatively inexpensive (but must be treated very carefully near edges), we can use Monte Carlo methods to generate large datasets of ZACVs for any assumed set of parameters describing their formation.  This allows us to explore the many dimensions of parameter space (e.g., CE efficiency, mass correlation) in an efficient manner.  We have also demonstrated the fidelity of this approach by taking our results (where applicable) and comparing them with previous studies.

This is the first time that a full CV population synthesis has been carried out for the present-day population that takes into account the effects of chemical evolution for all donor masses up to the bifurcation limit and for the complete spectrum of WD masses. The results show that the assumption that CV evolution can be approximated by ZAMS mass-losing donors is generally a valid one.  However, the range of the values of the observables at a given epoch has been greatly underestimated relative to other studies.  This is especially true for the relatively rare, long-period CVs ($P_{orb} > 6$ hr) where we find good agreement with the range of observed spectral types.  We are also able to populate the entire gap with non-magnetic CVs and predict the relative numbers that should be observed on both sides of the gap and in the gap itself.  We find that the relative distribution is not in contradiction with the observations, and we claim that a non-negligible fraction of systems in the gap may be derived from (partially) evolved donors.  We show that there should be a significant enhancement in the number of CVs observed near $P_{min}$.  If GR is the sole mechanism for AML below the period gap, then this increase in the detection probability should manifest itself in an 4 - 5 minute period range coincident with $P_{min}$. If we can obtain an unbiased and statistically significant sample of CVs whose orbital periods are close to the minimum value, the theoretical predictions of the population synthesis models could be used to determine if there is a supplemental AML mechanism acting contemporaneously and determine if its magnitude varies depending on the properties of the binary system (e.g., the chemical composition of the donor, the magnetic field strength of the WD, etc.).  With respect to ultracompacts, including AM CVn binaries, we conclude that the `CV channel' is not likely to account for any substantial fraction of these systems. In fact, if this channel has led to the formation of any of the observed ultracompacts, then it is very likely that they will be the ones with orbital periods of not much less than 60 minutes.

Two of the main difficulties encountered in using our PS approach concern the need to compute a suitable grid so that the interpolations are as accurate as possible and the numerical expense of computing a new grid if the physics governing the evolution of the donor and/or binary is changed.  With regard to the first issue, the grid needs to be carefully constructed so that difficult regions in initial-condition space that cause any of the parameters to change rapidly can be treated carefully (e.g., dynamical instabilities and the bifurcation limit).  We plan to add many additional tracks to our grid in an attempt to address this issue and to incorporate the contributions from brown dwarfs.  We will also extend the grid above the bifurcation limit and carry out a PS on these long-period systems that lead to mergers or the formation of double degenerates (including those containing hybrid HeCO WDs).  Finally, we plan to recalculate the grid by self-consistently incorporating the effects of mass gain (or erosion) of the WD accretor during the evolution.  Based on the very preliminary results presented in this paper, it appears as if WDs in PDCVs can gain $\sim 0.1 M_\odot$, an amount that may be consistent with the observations.

\acknowledgments

L. Nelson would like to thank the Canada Research Chairs Program and the Natural Sciences and Engineering Research Council (NSERC) of Canada for financial support. We also thank E. Dubeau, B. G\"{a}nsicke, S. Howell, C. Knigge, J. Patterson, D. Pawluczuk, S. Rappaport, and E. Sion for helpful discussions, and J. Hooey, A. Jacques, V. Kuczynski, and N. Roach for their technical assistance. We gratefully acknowledge the R\'eseau qu\'eb\'ecois de calcul de haute performance (RQCHP) and Calcul Qu\'ebec for providing the computational facilities required to carry out this work.  Finally, we would like to thank the referee for a large number of constructive comments and suggestions that greatly improved the paper.

\end{document}